\documentclass[appendixfloats]{emulateapj}
\usepackage{longtable}
\usepackage{graphicx}
\usepackage{graphics}
\usepackage{color}
\usepackage{epsfig}
\usepackage{rotating}
\usepackage{subfigure}
\usepackage{float}
\usepackage{url}
\usepackage{hyperref}
\usepackage{breakurl}
\usepackage{scalerel}
\usepackage[titletoc,title]{appendix}
\usepackage{mathptmx}
\usepackage{natbib}

\shortauthors{J. T. Li et al.}
\shorttitle{NuSTAR Observations of SN1006}

\begin{document}

\title{Spatially resolved broad-band synchrotron emission from the non-thermal limbs of SN1006}

\author{Jiang-Tao Li\altaffilmark{1}, Jean Ballet\altaffilmark{2}, Marco Miceli\altaffilmark{3,6}, Ping Zhou\altaffilmark{4}, Jacco Vink\altaffilmark{4}, Yang Chen\altaffilmark{5}, Fabio Acero\altaffilmark{2}, Anne Decourchelle\altaffilmark{2}, Joel N. Bregman\altaffilmark{1}} 

\altaffiltext{1}{Department of Astronomy, University of Michigan, 311 West Hall, 1085 S. University Ave, Ann Arbor, MI, 48109-1107, U.S.A.}

\altaffiltext{2}{Service d’Astrophysique, CEA Saclay, F-91191 Gif-sur-Yvette Cedex, France}

\altaffiltext{3}{Dipartimento di Fisica and Chimica, Universit\`a di Palermo, Piazza del Parlamento 1, I-90134 Palermo, Italy}

\altaffiltext{4}{Anton Pannekoek Institute/GRAPPA, University of Amsterdam, PO Box 94249, 1090 GE Amsterdam, The Netherlands}

\altaffiltext{5}{Department of Astronomy, Nanjing University, Nanjing 210023, China}

\altaffiltext{6}{INAF-Osservatorio Astronomico di Palermo, Piazza del Parlamento, I-90134 Palermo, Italy}

\keywords{acceleration of particles --- shock waves --- cosmic rays --- ISM: supernova remnants --- X-rays: ISM.}

\nonumber

\begin{abstract}
We present $\sim400\rm~ks$ \emph{NuSTAR} observations of the northeast (NE) and southwest (SW) non-thermal limbs of the Galactic SNR SN1006. We discovered three sources with X-ray emission detected at $\gtrsim50\rm~keV$. Two of them are identified as background AGN. We extract the \emph{NuSTAR} spectra from a few regions along the non-thermal limbs and jointly analyze them with the \emph{XMM-Newton} spectra and the radio data. The broad-band radio/X-ray spectra can be well described with a synchrotron emission model from a single population of CR electrons with a power law energy distribution and an exponential cutoff. The power law index of the electron particle distribution function (PDF) is $\approx1.88-1.95$ for both the NE and SW limbs, and we do not find significant evidence for a variation of this index at different energy (curvature). There are significant spatial variations of the synchrotron emission parameters. The highest energy electrons are accelerated in regions with the lowest expansion velocity, which is opposite to what has been found in the Tycho's SNR. In addition to a gradual steepening of synchrotron emission from the center of the non-thermal limbs to larger azimuthal angles, we also find that both the emission spectrum and the PDF are significantly flatter in three regions in the SW limb where the shock encounters higher density ambient medium. The NE limb also shows significantly higher cutoff energy in the PDF than the SW limb. By comparing with the roughly symmetric TeV emission and largely asymmetric GeV emission from the two non-thermal limbs, we conclude that the asymmetry in the ambient medium and magnetic fields may have largely modified the acceleration and emission of CR leptons.
\end{abstract}

\section{Introduction}\label{sec:Introduction}

Supernova remnants (SNRs) are thought to be efficient sites accelerating cosmic rays (CRs) up to the ``knee'' of Galactic CR spectrum (at $\sim3\times10^3\rm~TeV$; e.g., \citealt{Blandford87,Blasi13}). Non-thermal emissions in radio and hard X-rays associated with young SNRs are often best modeled with synchrotron emission of leptonic CRs, so have been adopted as direct evidence of shock acceleration of CR particles (e.g., \citealt{Koyama95}). The shape of the particle distribution function (PDF) responsible for the synchrotron emission is often characterized by two key parameters, the slope and cutoff energy. The slope of the PDF can be modified by non-linear acceleration processes and other micro-physics (e.g., \citealt{Caprioli12}), while the cutoff energy, or typically the maximum energy a shock accelerated particle can reach, could be limited by a few physical processes, such as the synchrotron radiative loss, the limited acceleration time, and the change of the available magnetohydrodynamic (MHD) waves above some wavelength (known as loss-, time-, or escape-limited accelerations; e.g., \citealt{Reynolds08}). 

The remnant of the supernova (SN) AD1006 (SN1006) is of particular scientific interest as a well known site of CR acceleration in a relatively clean environment (high Galactic latitude with no significant molecular cloud, e.g., \citealt{Dubner02}). The nonthermal nature of the X-ray spectrum was suggested as early as in the \emph{Einstein} era (e.g., \citealt{Becker80}), and was confirmed with \emph{ASCA} observations \citep{Koyama95}. In the two prominent limbs located in the northeast (NE) and southwest (SW) quarter of the SNR shell, the radio/X-ray correspondence is striking even in fine structures, suggesting that the nonthermal hard X-ray emission is primarily synchrotron radiation produced by the same population of shock-accelerated electrons responsible for the radio emission (e.g., \citealt{Winkler97}). Later \emph{Chandra} and \emph{XMM-Newton} observations resolved the fine structures of the nonthermal filaments (e.g., \citealt{Long03,Bamba03}), and revealed significant spatial variations of the synchrotron emission parameters (e.g., \citealt{Rothenflug04,Allen08,Miceli09,Miceli14a,Li15}).

Observing the spatial variation of the shape of multi-wavelength synchrotron spectrum provides us a lot of information about the physics of particle acceleration, such as the energy spectrum of the accelerated electrons, the strength and structure of the magnetic field, and the physical mechanisms limiting the maximum energy achieved by the accelerated electrons, etc. \citep{Reynolds08}. \citet{Dyer01} jointly fitted the available X-ray, radio, and TeV data and found that extrapolating from the measured radio fluxes grossly overpredicts the X-ray flux. \citet{Bamba08} make use of the wide-band \emph{Suzaku} observations to determine the roll-off frequency of the power law in X-rays, and found it is significantly higher in the NE than in the SW. \citet{Katsuda10} found that there is a spatial correlation between the nonthermal X-ray flux and the cutoff frequency in the NE limb. This result may indicate that the cutoff frequency depends on the magnetic-field strength, which is a natural prediction of the time- or escape-limited scenario, but not the loss-limited scenario \citep{Reynolds08}. However, \citet{Miceli13} found that a synchrotron radiative loss limited model provides better fit to all the X-ray spectra of the nonthermal limbs than acceleration time limited or escape limited models. These different conclusions can be reconciled in a loss limited scenario if the rate of particle injection and/or acceleration depends on some effects not yet accounted for, e.g., the shock obliquity.

Previous modeling of the broad-band synchrotron emission from radio to hard X-ray in SNRs is often limited by two issues: \textbf{(1)} the radio data either lacks spatial resolution in single-dish observations or is inaccurate in flux due to the limited $u, v$ coverage in interferometry observations. \textbf{(2)} the soft X-ray emission at $\lesssim(1-2)\rm~keV$ is often dominated by the thermal component and the hard X-ray emission at $\gtrsim(7-8)\rm~keV$ has low resolution and sensitivity.

In this paper, we present spatially resolved joint analyses of the broad-band synchrotron spectra from the two non-thermal limbs of SN1006, based on the high-resolution flux-accurate radio image from \citet{Dyer09}, the archival \emph{XMM-Newton} data analyzed in \citet{Li15,Li16}, and a large \emph{NuSTAR} program approved in Cycle~1. We will present analysis of the \emph{NuSTAR} data in \S\ref{sec:DataAnalysis}, joint analysis of the radio, \emph{XMM-Newton}, and \emph{NuSTAR} spectra in \S\ref{sec:results}, and discuss the results in \S\ref{sec:Discussions}. The conclusions will be summarized in \S\ref{section:Summary}. Errors are quoted at 1~$\sigma$ confidence level throughout the paper.


\section{NuSTAR Data Calibration and Analysis}\label{sec:DataAnalysis}


\subsection{NuSTAR Data Calibration}\label{subsec:DataCalibration}

\begin{figure}
\begin{center}
\epsfig{figure=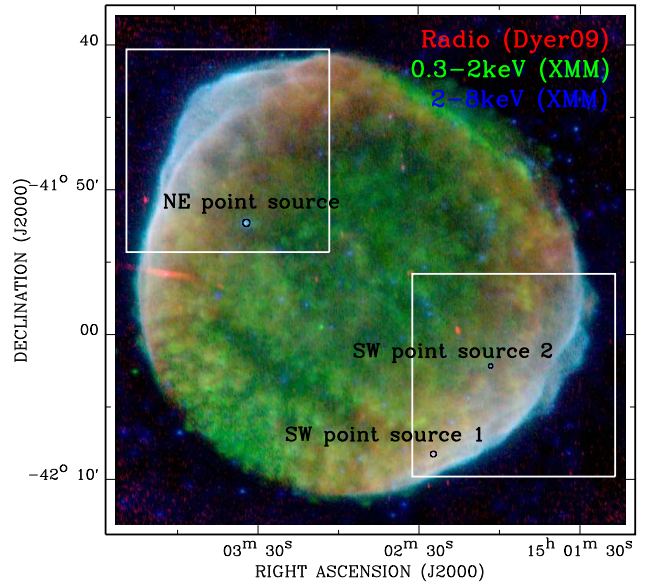,width=0.5\textwidth,angle=0, clip=}
\caption{Tri-color images of SN1006: red: radio images from \citet{Dyer09}; green and blue: 0.3-2~keV and 2-8~keV \emph{XMM-Newton} images from \citet{Li15}. The two white boxes roughly outline the FOV of the \emph{NuSTAR} observations. The small black circles are the \emph{XMM-Newton} spectral extraction regions of the three point sources detected by \emph{NuSTAR} (Figs.~\ref{fig:NuSTARtricolor}, \ref{fig:pointsrc}).}\label{fig:TriColorSN1006}
\end{center}
\end{figure}

\emph{NuSTAR} observations of the two non-thermal limbs of SN1006 are approved in Cycle~1 (2015-04-01 --- 2016-04-30; PI: Jiang-Tao Li) and are taken on 2016-03-02 [northeast (NE) limb; OBSID=40110001001 and 40110001002] and 2016-03-08 [southwest (SW) limb; OBSID=40110002001 and 40110002002], respectively, each lasting $\sim200\rm~ks$ (Fig.~\ref{fig:TriColorSN1006}).

We reprocess the \emph{NuSTAR} data using the standard \emph{NuSTAR} pipeline data reduction tool {\small nupipeline}. We have applied strict criteria regarding passages through the South Atlantic Anomaly (SAA) and a ``tentacle''-like region of higher activity near part of the SAA, i.e., by setting {\small SAAMODE=STRICT} and {\small TENTACLE=YES} when calling {\small nupipeline} to create Level~2 products. The background is stable with no strong flares in the lightcurve, so no further filtering has been adopted. The final effective exposure times are 181.49~ks (NE limb, telescope~A), 180.95~ks (NE limb, telescope~B), 188.37~ks (SW limb, telescope~A), and 187.63~ks (NE limb, telescope~B), respectively. Compared to a less strict setting of {\small SAAMODE=OPTIMIZED}, the resultant effective exposure time is $\sim7\%$ shorter.

\subsection{Background Analysis}\label{subsec:background}

We analyze the \emph{NuSTAR} background using the software package {\small nuskybgd} \citep{Wik14}. There are in general four major components of the \emph{NuSTAR} background, which are discussed in detail in \citet{Wik14}. \textbf{(1)} Internal background $I_d(E)$. The internal background is comprised of two components: a featureless continuum plus various activation and fluorescence lines. The lines dominate the background in 22-32~keV, although some weaker lines are still present at higher energy. There is no significant spatial variation of the internal background across a given detector. \textbf{(2)} Aperture Stray Light $A_d(E,x,y)$. This component is caused by the ``open'' design of the telescope, so a fraction of the unfocused X-rays could strike the detectors. It is therefore expected that this ``Aperture'' background depends on the position (x,y) on the focal plane. The spectral shape, on the other hand, is consistent with the cosmic X-ray background (CXB). \textbf{(3)} Scattered and Reflected Stray Light $S_d(E)$. This component is caused by the scattered and reflected X-rays from the entire sky (the CXB, the Earth, and the Sun) by different parts of the spacecraft (e.g., the mast). The scattered or reflected CXB emission could be directly added to the $A_d(E,x,y)$ component, while those from the Earth and the Sun are much softer and can be described with a thermal component ($\sim1\rm~keV$; typically dominated below $\sim5\rm~keV$). Some weak fluorescence lines can be added to the $I_d(E)$ term. Therefore, the primary component of this $S_d(E)$ background is made up of the ``Solar'' emission and has no spatial variation. \textbf{(4)} Focused Cosmic Background $f_d(E,x,y)$. Like in other X-ray missions, the focused cosmic background (fCXB) is produced by unresolved foreground/background sources. This component is in general less important for \emph{NuSTAR} and can only be noticeable below 15~keV. The fCXB background has some spatial variations, which depends on the direction of the observations.

\begin{figure*}
\begin{center}
\epsfig{figure=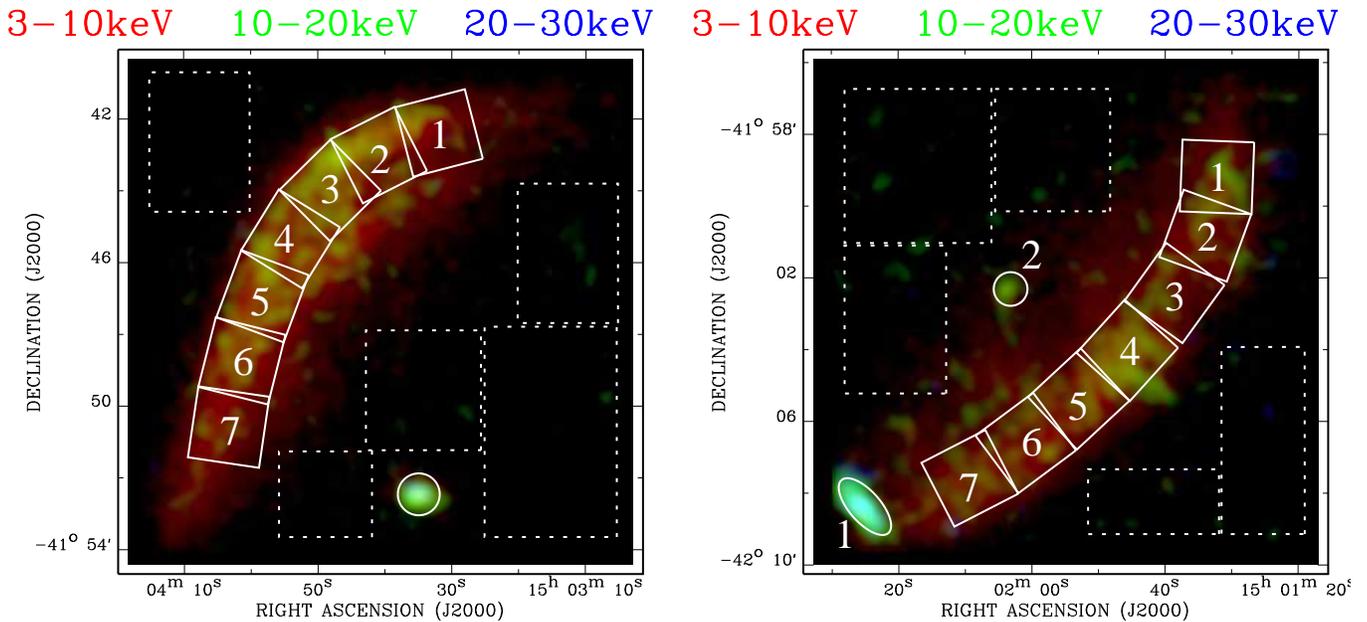,width=1.0\textwidth,angle=0, clip=}
\caption{\emph{NuSTAR} tri-color (red: 3-10~keV; green: 10-20~keV; blue: 20-30~keV) images of the NE and SW limbs from the two white boxes in Fig.~\ref{fig:TriColorSN1006}. Solid regions (boxes, circle, and ellipse) are used to extract source spectra, while dashed boxes are those used to extract background spectra. The box regions used to extract spectra from the two non-thermal limbs are labeled ``1-7''. The two point-like sources in the FOV of the SW limb are also labeled ``1'' and ``2''.}\label{fig:NuSTARtricolor}
\end{center}
\end{figure*}

\begin{figure*}
\begin{center}
\epsfig{figure=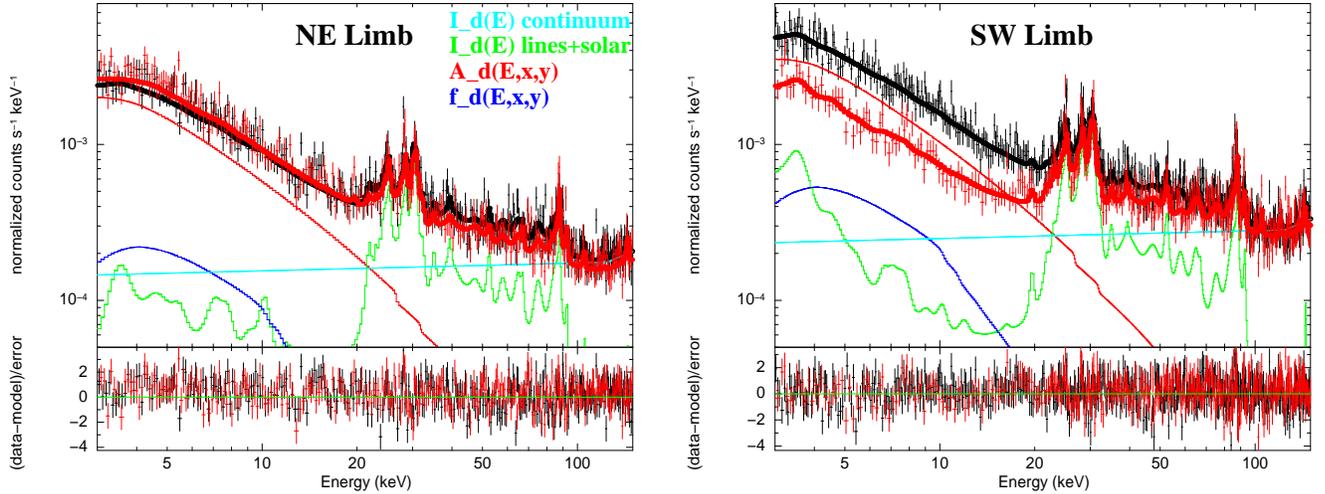,width=1.0\textwidth,angle=0, clip=}
\caption{Background spectra of the NE (left) and SW (right) FOV. Black and red data points represent spectra extracted from telescopes A and B respectively, but we just show an example of one of the five background regions in each FOV (dotted boxes in Fig.~\ref{fig:NuSTARtricolor}). The spectra extracted from different regions, however, are jointly fitted, in order to sample the spatial variation of different background components. There are basically four background components: Internal $I_d(E)$ (two component: a featureless continuum plus lines), Aperture Stray Light $A_d(E,x,y)$, Scattered and Reflected Stray Light $S_d(E)$ (``solar'' and part of $A_d$), and Focused Cosmic Background $f_d(E,x,y)$ \citep{Wik14}, which are plotted with thin curves in different colors. Background components with spatial variation are denoted with $(x,y)$. The thick solid curve is the sum of different background components (best-fit).}\label{fig:bgdspec}
\end{center}
\end{figure*}

For background analysis using {\small nuskybgd}, we first define a few (five) background regions (dashed boxes in Fig.~\ref{fig:NuSTARtricolor}) for each field of view (FOV; for the NE and SW limbs), in order to sample the spatial variation of different background components. These background regions are known to have negligible non-thermal X-ray emission, although thermal emission typically $\lesssim2\rm~keV$ does present in the SNR interior (Fig.~\ref{fig:TriColorSN1006}, e.g., \citealt{Li15,Li16}). We then make instrument maps using the {\small IDL} tool {\small nuskybgd\_instrmap} and create detector and background aperture images using {\small projinitbgds}. The background spectra are extracted using the {\small PYTHON} tool {\small getspecnoarf}. We then jointly fit the background spectra extracted from different regions using {\small nuskybgd\_fitab}; this tool simultaneously create background parameter files which will be used to create background images (using {\small nuskybgd\_image}) and background spectra for a source region (\S\ref{subsec:spec}). The above background analyses are conducted for the NE and SW FOVs separately. Examples of the fitted background spectra in the NE and SW FOVs are presented in Fig.~\ref{fig:bgdspec}.

\subsection{Spectral Analysis}\label{subsec:spec}

For \emph{NuSTAR} spectral analysis of extended sources, local sky backgrounds may not be accurate enough as they do not account for the spatial variation of some background components (\S\ref{subsec:background}). We therefore create simulated background spectra for each source region based on the background spectral parameters obtained in \S\ref{subsec:background}. This is achieved using the {\small nuskybgd} tool {\small nuskybgd\_spec}. We also extract source spectra using the standard {\small HEASOFT} tool {\small nuproducts}. 

The above spectra extraction is adopted for a few source regions as presented in Fig.~\ref{fig:NuSTARtricolor}. Each source region is a $2^\prime\times2^\prime$ box roughly along the non-thermal limbs. The size of the boxes is significantly larger than the FWHM (Full Width at Half Maximum; $18^{\prime\prime}$) or the HPD (Half Power Diameter; $58^{\prime\prime}$) of \emph{NuSTAR} \citep{Harrison13}. Therefore, the spectra are not significantly affected by the spilling out photons in the PSF wing or the scattered light from surrounding regions. Below we will discuss the modeling of the spectra extracted from various regions along the non-thermal limbs. The spectra of a few bright point-like sources will be presented in \S\ref{subsec:PointSource}.

We jointly fit the \emph{NuSTAR} spectra from this work, the \emph{XMM-Newton} spectra from \citet{Li15,Li16}, and the 1.37~GHz radio flux measured from the flux-accurate image constructed with both the \emph{VLA} and \emph{GBT} data \citep{Dyer09} with the Interactive Spectral Interpretation System (ISIS; \citealt{Houck00}). The spectra or flux in different bands are extracted from the same regions, which are significantly larger than the PSF of each observations. These multi-wavelength spectra are fitted with various non-thermal emission models, with all the physical parameters linked and a constant scaling factor multiplied to each X-ray spectrum in order to account for small uncertainties in the normalization (caused by the bad pixels, CCD gaps, or the empty regions out of the FOV). We fix this constant at 1 for the \emph{NuSTAR} telescope~A and the radio flux, in order to make sure that the joint flux calibration is accurate and not biased by the scaling factor. The scaling factors of \emph{NuSTAR} telescope~B and the \emph{XMM-Newton} spectra extracted from different instruments (MOS-1, MOS-2, PN) and  different observations are all allowed to vary. The best-fit constants of different spectra are in general close to 1, ensuring the accuracy of the cross flux calibration of different data.

\subsection{Point-like Sources}\label{subsec:PointSource}

There are three prominent hard X-ray bright point-like sources detected by the two \emph{NuSTAR} observations, one in the NE FOV, the other two in the SW FOV (Fig.~\ref{fig:NuSTARtricolor}). The apparently elongated morphology of the SW point source~1 is indeed consistent with the shape of the PSF at the corresponding location \citep{An14}. All these three sources have soft X-ray counterparts detected by \emph{XMM-Newton}, but only the one located in the NE FOV is significantly brighter than the soft X-ray knots of the SN ejecta (Fig.~\ref{fig:TriColorSN1006}). We cross identify these three sources with online catalogues (such as SIMBAD, \citealt{Wenger00}). Two sources, the NE source and the SW source~1, have multi-wavelength counterparts and are both identified as background AGN (Table~\ref{table:pointsrc}). The third source, SW source~2, although seems to have a very faint optical counterpart, is not identified in other catalogues.

\begin{figure*}
\begin{center}
\epsfig{figure=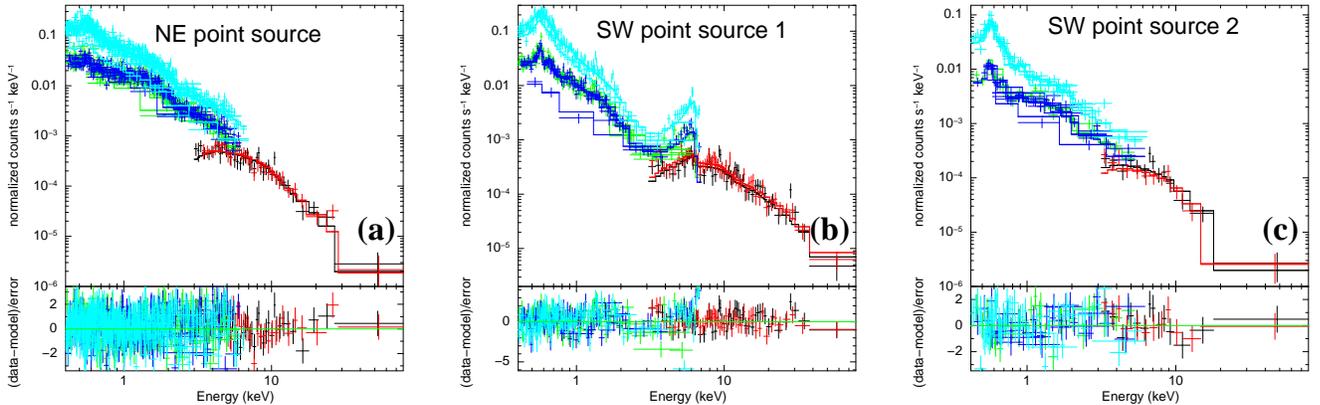,width=1.0\textwidth,angle=0, clip=}
\caption{\emph{NuSTAR} spectra (black: telescope A; red: telescope B) of the three most luminous point-like sources in the FOV (Fig.~\ref{fig:NuSTARtricolor}). The spectra are fitted with a power law (a,c) or broken power (b) plus a thermal plasma (NEI) and background components. The best-fit photon indexes of each source are also marked in the corresponding panel. An additional broadened Fe~K line from the accretion disk of a supermassive black hole (kerrdisk) is also needed to fit the spectra of the SW point source 1 (b).
}\label{fig:pointsrc}
\end{center}
\end{figure*}


\begin{table}
\begin{center}
\caption{Parameters of the point-like sources detected by \emph{NuSTAR}} 
\footnotesize
\tabcolsep=1.8pt%
\begin{tabular}{lccccccccccccc}
\hline
Name & NE & SW 1 & SW 2 \\
\hline
RA & 15h03m34.1s & 15h02m24.9s & 15h02m03.5s \\
DEC & -41d52m24.1s & -42d08m24.4s & -42d02m19.9s \\
identifier & QSOJ1503-4152 & 2MASXJ15022467-4208244 & - \\
type & quasar & galaxy & - \\
redshift & 0.335 & 0.05388 & - \\
$d$ ($^{\prime\prime}$) & 1.23 & 2.46 & - \\
model & power+NEI & bknpower+kerrdisk+NEI & power+NEI \\
$\Gamma_1$ & $1.65\pm0.02$ & $2.58_{-0.04}^{+0.03}$ & $1.82\pm0.11$ \\
$\Gamma_2$ & - & $0.57\pm0.02$ & - \\
$E_{\rm break}$ & - & $3.21_{-0.05}^{+0.04}$ & - \\
$spin_{\rm XMM}$ & - & $0.024_{-0.002}^{+0.001}$ & - \\
$F_{\rm 0.5-3 keV}$ & $0.48\pm0.02$ & $0.52\pm0.01$ & $0.16\pm0.02$ \\
$F_{\rm 3-10 keV}$ & $1.27\pm0.06$ & $1.99\pm0.06$ & $0.40\pm0.03$ \\
$F_{\rm 10-30 keV}$ & $1.73\pm0.08$ & $9.06\pm0.28$ & $0.45\pm0.06$ \\
$F_{\rm 30-80 keV}$ & $2.22\pm0.12$ & $35.21_{-1.90}^{+1.08}$ & $0.49_{-0.10}^{+0.12}$ \\
$F_{\rm kerr,NuSTAR}$ & - & $0.20\pm0.05$ & - \\
$F_{\rm kerr,XMM}$ & - & $0.43_{-0.01}^{+0.03}$ & - \\
\hline
\end{tabular}\label{table:pointsrc}
\end{center}
The sources are cross-identified with SIMBAD \citep{Wenger00}, with the closest identifier having a distance of $d$ from the \emph{NuSTAR} source. $\Gamma_1$ and $\Gamma_2$ are the photon indexes of the power law or broken power law component. $E_{\rm break}$ is the break energy of the broken power law. Fluxes in different bands (in unit of $10^{-13}\rm~ergs~s^{-1}~cm^{-2}$) are measured only for the power law or broken power law component after removing the NEI component representing the plasma from the SNR and the background components. For the SW point source~1, we add another component (``kerrdisk''; \citealt{Brenneman06}) describing the broad Fe~K line. As this line has significantly different strength in the \emph{XMM-Newton} and \emph{NuSTAR} spectra, we allow the parameters of this component to be different between the two sets of spectra. Since the Fe~K line in the \emph{NuSTAR} spectra is too weak, we only list the dimensionless black hole spin determined with the \emph{XMM-Newton} data ($spin_{\rm XMM}$). $F_{\rm kerr,NuSTAR}$ and $F_{\rm kerr,XMM}$ are the 3-10~keV fluxes (also in unit of $10^{-13}\rm~ergs~s^{-1}~cm^{-2}$) of the kerrdisk component of SW source~1 measured with the \emph{NuSTAR} and \emph{XMM-Newton} data, respectively.
\end{table}

These three sources all have harder X-ray spectra than the non-thermal limbs, so we have a few data points above $30\rm~keV$ which are not too strongly affected by the instrumental lines (Fig.~\ref{fig:pointsrc}). We extract \emph{XMM-Newton} spectra from smaller regions than the \emph{NuSTAR} spectral extraction regions (Fig.~\ref{fig:TriColorSN1006}), in order to avoid too much contribution from the thermal plasma in the surrounding area, which is not important in the \emph{NuSTAR} band. We then need to multiply a scaling factor to account for this difference in area scale when jointly analyzing the \emph{XMM-Newton} and \emph{NuSTAR} spectra.

Selected parameters of the X-ray spectral models are summarized in Table~\ref{table:pointsrc}. The \emph{XMM-Newton} and \emph{NuSTAR} spectra of the NE point source and SW point source~2 can both be well fitted with a single power law plus thermal (described with a NEI model) and background components (Fig.~\ref{fig:pointsrc}a,c). The power law photon index of the SW point source~2 is consistent with those of X-ray binaries, and since it does not have a bright optical or IR counterpart, it is most likely a Galactic source. The SW point source~1 has much more complicated X-ray spectra which include an additional Fe~K line with a broadened low energy wing (Fig.~\ref{fig:pointsrc}b). This feature can be roughly fitted with a model describing the emission from the accretion disk of a supermassive black hole (``kerrdisk''; \citealt{Brenneman06}). Furthermore, the broad-band non-thermal continuum cannot be fitted with a single power law, and a significant flattening of the spectra at $\gtrsim3.2\rm~keV$ is clearly revealed by \emph{NuSTAR}. This source has a near-IR counterpart which is a background galaxy (Table~\ref{table:pointsrc}), so most likely to be the AGN of this galaxy. 


\section{Results from spectral analysis}\label{sec:results}


\subsection{Curvature of synchrotron X-ray spectra}\label{subsec:XrayCurv}

\begin{figure*}
\begin{center}
\epsfig{figure=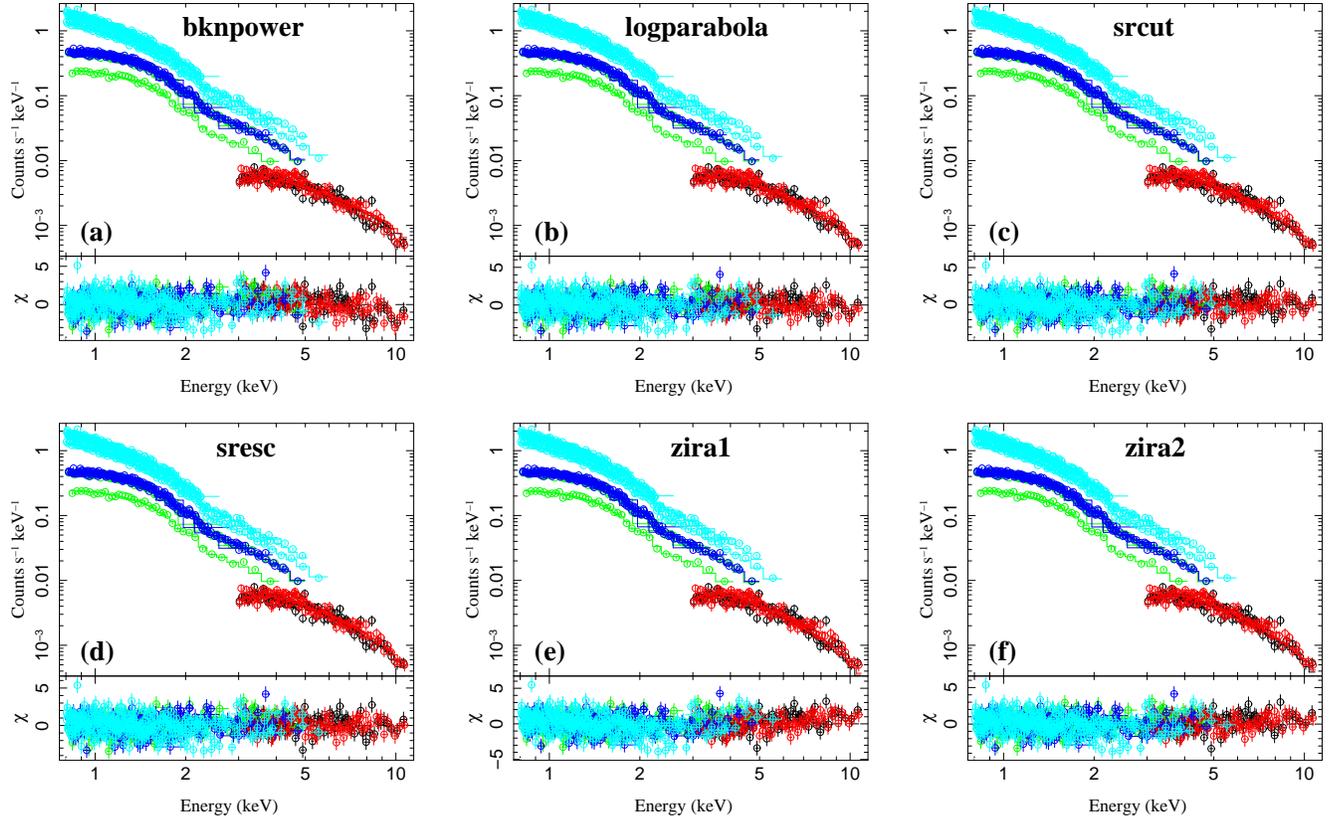,width=1.0\textwidth,angle=0, clip=}
\caption{Example \emph{NuSTAR} (black: telescope A; red: telescope B) and \emph{XMM-Newton} (green: MOS-1; blue: MOS-2; cyan: PN; only data at $>0.8\rm~keV$ are included in the fit and plot) spectra of a source region ``3'' in the NE Limb (Fig.~\ref{fig:NuSTARtricolor}). Similar plots of other regions are included in the appendix (Fig.~\ref{fig:specotherreg}). The spectra are fitted with different models: (a) broken power law; (b) log-parabolic; (c) srcut; (d) sresc; (e) zira1; (f) zira2, as described in the text. All these models subject to foreground absorption (described with a ``wabs'' model) with the column density fixed at the direction of SN1006 ($N_{\rm H}=6.8\times10^{20}\rm~cm^{-2}$; \citealt{Dubner02}). Each spectrum (except for \emph{NuSTAR} telescope~A) is also renormalized by a scaling factor close to 1 in order to account for the small calibration bias. Best-fit parameters of different models are summarized in Tables~\ref{table:NuSTARXMMpowerpara} - \ref{table:zirapara}.}\label{fig:specXray}
\end{center}
\end{figure*}

We first use a few mathematical models to characterize the overall shape of the broad-band non-thermal X-ray continuum from 0.8~keV to $>10\rm~keV$ (panels a,b of Figs.~\ref{fig:specXray} and \ref{fig:specotherreg}). We characterize the shape of the non-thermal spectra in three ways: \textbf{(1)} use two different power law models to fit the \emph{NuSTAR} and \emph{XMM-Newton} spectra, respectively; use \textbf{(2)} a broken power law (``bknpower'') or \textbf{(3)} a curved power law model described with the fomula $f(x)=norm*(x/x_{\rm ref})^{-\gamma-\beta*\log_{10}(x/x_{\rm ref})}$ (``logparabola'', where $x_{\rm ref}$ is fixed at 1~keV; \citealt{Massaro04}) to jointly fit the \emph{NuSTAR} and \emph{XMM-Newton} spectra simultaneously.

We fit the spectra extracted from all the box regions in Fig.~\ref{fig:NuSTARtricolor} with these models and present the fitted spectra with the bknpower and log-parabolic models in panels (a,b) of Figs.~\ref{fig:specXray} and \ref{fig:specotherreg}. The best-fit physical parameters, their 1~$\sigma$ errors, the reduced $\chi^2$, and degree of freedom (d.o.f.) of each region are summarized in Tables~\ref{table:NuSTARXMMpowerpara} (for the power law model fitting the \emph{NuSTAR} and \emph{XMM-Newton} spectra, respectively), \ref{table:Xdatabknpowerpara} (for the bknpower model), and \ref{table:Xdatalogparabolapara} (for the log-parabolic model). In all these spectral analysis, we have adopted foreground absorption column density toward SN1006 ($N_{\rm H}$ fixed at $6.8\times10^{20}\rm~cm^{-2}$; \citealt{Dubner02}). The exact value of $N_{\rm H}$ does not affect the non-thermal X-ray emission at $\gtrsim0.8\rm~keV$ significantly. We removed the data $\gtrsim20\rm~keV$ in the \emph{NuSTAR} spectra, where the synchrotron emission is too weak and the residual background from some strong emission lines at $\sim20-30\rm~keV$ (Fig.~\ref{fig:bgdspec}) make the spectra much more uncertain than those at lower energy. We also remove the data points at $\lesssim0.8\rm~keV$ in the \emph{XMM-Newton} spectra, where the strong oxygen emission lines from a thermal plasma component (much stronger than those in Fig.~\ref{fig:pointsrc}) may affect the fitting of the pure non-thermal emission. We add the sky background models described in \citet{Li15} to the \emph{XMM-Newton} spectra, which are extracted from an annulus out of the SNR and have been rescaled to each regions according to their effective sky area.


\begin{table}
\begin{center}
\caption{Best-fit power law index in \emph{NuSTAR} and \emph{XMM-Newton} bands} 
\footnotesize
\tabcolsep=3.2pt%
\begin{tabular}{lccrccccccccccc}
\hline
Region & $\Gamma_{\rm NuSTAR}$ & $\chi^2/\rm d.o.f.$ & d.o.f $\mid$ & $\Gamma_{\rm XMM}$ & $\chi^2/\rm d.o.f.$ & d.o.f \\
\hline
NE\_1 & $3.276_{-0.102}^{+0.105}$ & 1.116 &   91 $\mid$ & $2.677\pm0.007$ & 1.932 & 3002 \\
NE\_2 & $3.011_{-0.067}^{+0.069}$ & 0.949 &  157 $\mid$ & $2.520\pm0.006$ & 1.643 & 3446 \\
NE\_3 & $2.948_{-0.064}^{+0.065}$ & 1.140 &  175 $\mid$ & $2.495\pm0.006$ & 1.551 & 3524 \\
NE\_4 & $3.099_{-0.068}^{+0.070}$ & 0.865 &  162 $\mid$ & $2.528\pm0.006$ & 1.492 & 3351 \\
NE\_5 & $2.955_{-0.065}^{+0.067}$ & 1.174 &  170 $\mid$ & $2.503\pm0.006$ & 1.458 & 3627 \\
NE\_6 & $3.211_{-0.074}^{+0.077}$ & 0.843 &  144 $\mid$ & $2.562\pm0.006$ & 1.533 & 3851 \\
NE\_7 & $3.372_{-0.101}^{+0.105}$ & 1.096 &  100 $\mid$ & $2.675\pm0.005$ & 1.893 & 5043 \\
SW\_1 & $2.978_{-0.118}^{+0.122}$ & 0.708 &   66 $\mid$ & $2.703\pm0.008$ & 2.069 & 3069 \\
SW\_2 & $3.208_{-0.098}^{+0.102}$ & 0.815 &   84 $\mid$ & $2.646\pm0.008$ & 2.202 & 2879 \\
SW\_3 & $3.134_{-0.108}^{+0.112}$ & 1.264 &   79 $\mid$ & $2.701\pm0.010$ & 2.105 & 2110 \\
SW\_4 & $3.043_{-0.074}^{+0.077}$ & 0.815 &  136 $\mid$ & $2.664\pm0.009$ & 1.610 & 1720 \\
SW\_5 & $3.158_{-0.075}^{+0.078}$ & 0.955 &  126 $\mid$ & $2.690\pm0.009$ & 1.380 & 1585 \\
SW\_6 & $3.159_{-0.078}^{+0.081}$ & 0.822 &  114 $\mid$ & $2.694\pm0.008$ & 1.323 & 1705 \\
SW\_7 & $3.118_{-0.087}^{+0.090}$ & 1.065 &  100 $\mid$ & $2.702\pm0.008$ & 1.326 & 1728 \\
\hline
\end{tabular}\label{table:NuSTARXMMpowerpara}
\end{center}
Spectral extraction regions are presented in Fig.~\ref{fig:NuSTARtricolor}. $\Gamma_{\rm NuSTAR}$ and $\Gamma_{\rm XMM}$ are the photon indexes obtained by fitting only the \emph{NuSTAR} and \emph{XMM-Newton} spectra, respectively. All the spectra have been rebinned to a signal-to-noise ratio of $\rm SNR=5$. The spectra are fitted after subtracting various background components (\S\ref{subsec:background}). Errors are quoted at 1~$\sigma$ level.
\end{table}

In the joint analyses of the spectra with significantly different counting statistics, such as the \emph{XMM-Newton} and \emph{NuSTAR} spectra of SN1006, the $\chi^2/\rm d.o.f.$ may not accurately reflect the goodness of the fit in a certain band. As we are more interested in the high energy band poorly constrained in previous observations (typically at $\gtrsim7\rm~keV$), we justify the goodness of the model not only by the $\chi^2/\rm d.o.f.$ for the whole spectra, but also by the presence of any systematical departure of the data from the ``best-fit'' model. Furthermore, we also regroup the \emph{NuSTAR} and \emph{XMM-Newton} spectra with different signal-to-noise ratios ($\rm S/N=5$ for \emph{NuSTAR} and $\rm S/N=20$ for \emph{XMM-Newton}), so that there will not be too many \emph{XMM-Newton} data points compared to the relatively few \emph{NuSTAR} data points. The goodness of the fit in the \emph{NuSTAR} band could then be reflected by the relatively small change of the $\chi^2/\rm d.o.f.$ in the fitting process. 


\begin{figure}
\begin{center}
\epsfig{figure=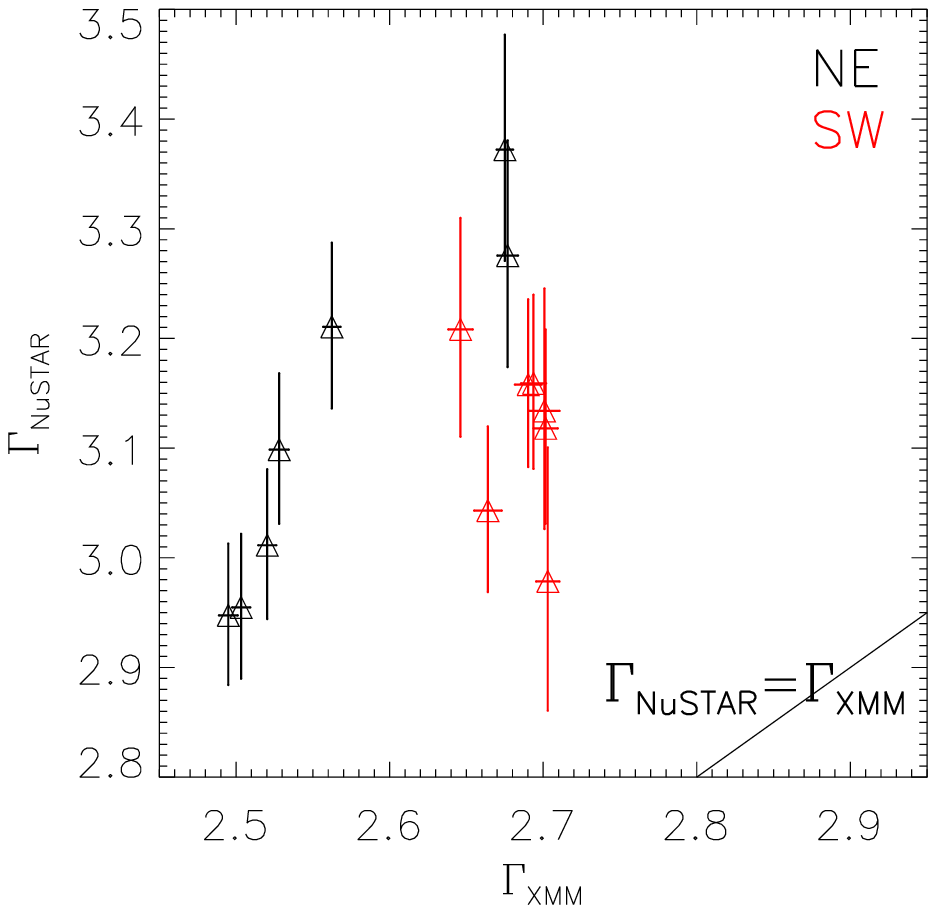,width=0.48\textwidth,angle=0, clip=}
\caption{The photon indexes of the power law fit to only the \emph{XMM-Newton} ($x$-axis) or \emph{NuSTAR} data ($y$-axis). Black and red data points are measurements from the NE and SW limb, respectively. The solid line is where $\Gamma_{\rm NuSTAR}=\Gamma_{\rm XMM}$. It is obvious that the spectra in the \emph{NuSTAR} band are systematically steeper than those in the \emph{XMM-Newton} band.}\label{fig:PhoIndexXMMNuSTAR}
\end{center}
\end{figure}

The spectral slope in the \emph{NuSTAR} band ($\Gamma_{\rm NuSTAR}\sim2.9-3.4$; typically from 3~keV to $\sim15\rm~keV$) is systematically steeper than those in the \emph{XMM-Newton} band ($\Gamma_{\rm XMM}\sim2.5-2.7$; typically 0.8-7~keV) (Table~\ref{table:NuSTARXMMpowerpara}). This systematical bias indicates a significant curvature of the broad-band non-thermal X-ray emission spectra. There seems to be a positive correlation between $\Gamma_{\rm NuSTAR}$ and $\Gamma_{\rm XMM}$ for the NE limb, but all regions in the SW limb are compatible with the same spectral slope in both \emph{NuSTAR} and \emph{XMM-Newton} bands (Fig.~\ref{fig:PhoIndexXMMNuSTAR}).


In most of the cases, the log-parabolic model fit the broad-band X-ray spectra better than a bknpower model (Tables~\ref{table:Xdatabknpowerpara}, \ref{table:Xdatalogparabolapara}, Fig.~\ref{fig:reducedchiAZ}a), with the bknpower model tends to overpredict the X-ray emission at $\gtrsim8\rm~keV$ (e.g., Fig.~\ref{fig:specXray}a). This indicates that the broad-band synchrotron emission is curved and more complicated than a broken power law with a single break.


\begin{table}
\begin{center}
\caption{Best-fit parameters of the bknpower model} 
\footnotesize
\tabcolsep=4.5pt%
\begin{tabular}{lcccccccccccccc}
\hline
Region & $\Gamma_1$ & $E_{\rm break}$/keV & $\Gamma_2$ & $\chi^2/\rm d.o.f.$ & d.o.f \\
\hline
NE\_1 & $2.467_{-0.023}^{+0.021}$ & $1.657_{-0.087}^{+0.084}$ & $2.834_{-0.026}^{+0.028}$ & 1.777 &  706 \\
NE\_2 & $2.389_{-0.034}^{+0.024}$ & $2.093_{-0.335}^{+0.370}$ & $2.705_{-0.063}^{+0.087}$ & 1.642 &  873 \\
NE\_3 & $2.349_{-0.019}^{+0.017}$ & $1.853_{-0.121}^{+0.137}$ & $2.641_{-0.027}^{+0.031}$ & 1.583 &  854 \\
NE\_4 & $2.375\pm0.015$ & $1.921_{-0.093}^{+0.097}$ & $2.719_{-0.026}^{+0.029}$ & 1.375 &  826 \\
NE\_5 & $2.359_{-0.014}^{+0.013}$ & $2.064_{-0.109}^{+0.122}$ & $2.711_{-0.031}^{+0.037}$ & 1.455 &  877 \\
NE\_6 & $2.360_{-0.025}^{+0.024}$ & $1.570_{-0.084}^{+0.097}$ & $2.673_{-0.020}^{+0.022}$ & 1.500 &  919 \\
NE\_7 & $2.515_{-0.011}^{+0.012}$ & $1.890_{-0.056}^{+0.073}$ & $2.861_{-0.023}^{+0.026}$ & 1.722 & 1053 \\
SW\_1 & $2.449_{-0.029}^{+0.030}$ & $1.408_{-0.053}^{+0.070}$ & $2.802_{-0.022}^{+0.025}$ & 1.795 &  614 \\
SW\_2 & $2.373\pm0.035$ & $1.436_{-0.066}^{+0.096}$ & $2.763_{-0.027}^{+0.031}$ & 2.005 &  566 \\
SW\_3 & $2.414_{-0.039}^{+0.077}$ & $1.438_{-0.069}^{+0.252}$ & $2.830_{-0.033}^{+0.099}$ & 1.904 &  400 \\
SW\_4 & $2.551_{-0.040}^{+0.016}$ & $2.327_{-0.469}^{+0.204}$ & $2.947_{-0.121}^{+0.067}$ & 1.337 &  474 \\
SW\_5 & $2.518\pm0.020$ & $1.829_{-0.083}^{+0.099}$ & $2.896_{-0.031}^{+0.034}$ & 1.274 &  480 \\
SW\_6 & $2.577\pm0.020$ & $2.166_{-0.209}^{+0.259}$ & $2.942_{-0.055}^{+0.076}$ & 1.192 &  514 \\
SW\_7 & $2.566_{-0.018}^{+0.017}$ & $1.901_{-0.085}^{+0.128}$ & $2.871_{-0.028}^{+0.033}$ & 1.273 &  513 \\
\hline
\end{tabular}\label{table:Xdatabknpowerpara}
\end{center}
Parameters of the broken power law model (bknpower) obtained by jointly fitting the \emph{NuSTAR} and \emph{XMM-Newton} spectra (panel~a of Figs.~\ref{fig:specXray} and \ref{fig:specotherreg}). The \emph{NuSTAR} spectra have been rebinned to a signal-to-noise ratio of $\rm SNR=5$, while the \emph{XMM-Newton} spectra have been rebinned to a signal-to-noise ratio of $\rm SNR=20$. $\Gamma_1$ and $\Gamma_2$ are the photon indexes below and above the break energy $E_{\rm break}$.
\end{table}


\begin{table}
\begin{center}
\caption{Best-fit parameters of the logparabola model} 
\footnotesize
\tabcolsep=5.0pt%
\begin{tabular}{lcccccccccccccc}
\hline
Region & $\gamma$ & $\beta$ & $\chi^2/\rm d.o.f.$ & d.o.f \\
\hline
NE\_1 & $2.408\pm0.017$ & $0.521\pm0.038$ & 1.657 &  707 \\
NE\_2 & $2.304\pm0.015$ & $0.378_{-0.028}^{+0.029}$ & 1.524 &  874 \\
NE\_3 & $2.285\pm0.016$ & $0.377_{-0.029}^{+0.030}$ & 1.479 &  855 \\
NE\_4 & $2.296\pm0.016$ & $0.431_{-0.030}^{+0.032}$ & 1.267 &  827 \\
NE\_5 & $2.275\pm0.015$ & $0.406_{-0.029}^{+0.030}$ & 1.386 &  878 \\
NE\_6 & $2.325\pm0.015$ & $0.438_{-0.029}^{+0.030}$ & 1.361 &  920 \\
NE\_7 & $2.439\pm0.014$ & $0.442_{-0.030}^{+0.031}$ & 1.681 & 1054 \\
SW\_1 & $2.460\pm0.020$ & $0.502_{-0.048}^{+0.051}$ & 1.800 &  615 \\
SW\_2 & $2.368\pm0.021$ & $0.572_{-0.048}^{+0.051}$ & 1.893 &  567 \\
SW\_3 & $2.419\pm0.025$ & $0.573_{-0.059}^{+0.063}$ & 1.836 &  401 \\
SW\_4 & $2.455\pm0.021$ & $0.392_{-0.042}^{+0.043}$ & 1.249 &  475 \\
SW\_5 & $2.458\pm0.021$ & $0.444_{-0.041}^{+0.042}$ & 1.197 &  481 \\
SW\_6 & $2.489\pm0.020$ & $0.398\pm0.039$ & 1.101 &  515 \\
SW\_7 & $2.504\pm0.019$ & $0.371_{-0.038}^{+0.039}$ & 1.207 &  514 \\
\hline
\end{tabular}\label{table:Xdatalogparabolapara}
\end{center}
Similar as Table~\ref{table:Xdatabknpowerpara}, but for the parameters of the log-parabolic model obtained by jointly fitting the \emph{NuSTAR} and \emph{XMM-Newton} spectra (panel~b of Figs.~\ref{fig:specXray} and \ref{fig:specotherreg}). $\gamma$ is the power law index, while $\beta$ is the curvature. The reference energy $x_{\rm ref}$ is fixed at 1~keV (\S\ref{subsec:XrayCurv}).
\end{table}


\subsection{Loss- and escape-limited models in X-ray band}\label{subsec:LossEscapeXray}

We next examine some physical synchrotron emission models in the fitting of the broad-band X-ray spectra, assuming all the synchrotron emissions are produced by a single population of CR electrons. We have adopted four loss- or escape-limited models all with two free parameters: \textbf{(1,2)} The synchrotron spectrum described in \citet{Reynolds98}, from either an exponentially cutoff power law distribution of electrons in a homogeneous magnetic field (``srcut''), or an electron distribution limited by particle escape above a break energy (``sresc''). We caution that these two models may be oversimplied with the homogeneous magnetic field assumption, as the spatial variation of the magnetic field in the upstream may change the shape of the synchrotron spectrum in the cutoff region. \textbf{(3,4)} The loss-limited models described in \citet{Zirakashvili07}. We adopt the two analytical forms described in \citet{Miceli13}, with slightly different parameters depending on whether the magnetic field downstream is compressed by a factor $\kappa=\sqrt{11}$ (``zira2'') or not (``zira1'') with respect to upstream. We have fixed the normalization of the srcut and sresc models at the radio flux extracted from the flux accurate image of \citet{Dyer09}, so all these four models have just two free physical parameters (plus a constant scaling factor of each spectra which is $\approx 1$, except for \emph{NuSTAR} telescope~A): the radio spectral index $\alpha$ and cutoff frequency $\nu_{\rm cutoff}$ (converted to energy unit $E_{\rm cutoff}$ for the convenience of comparison) of srcut and sresc, and the cutoff energy ($E_{\rm cut}$) and normalization of zira1 and zira2. The best-fit spectra are presented in panels (c-f) of Figs.~\ref{fig:specXray} and \ref{fig:specotherreg}, while the model parameters are listed in Tables~\ref{table:srcutsrescpara} and \ref{table:zirapara}.


\begin{figure*}
\begin{center}
\epsfig{figure=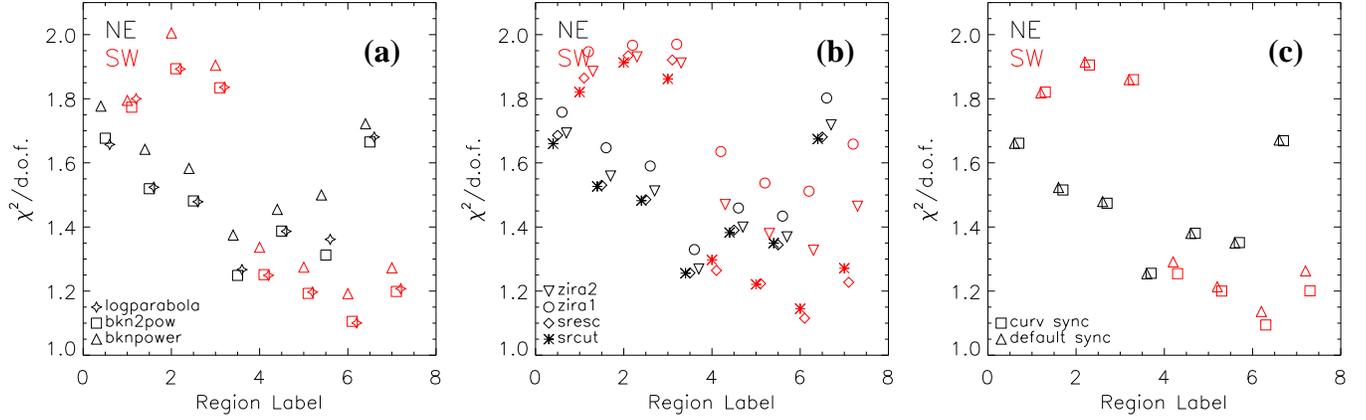,width=1.0\textwidth,angle=0, clip=}
\caption{The reduced $\chi^2$ of the fitted spectra of different regions. Different symbols represent reasonable fit with different models, as denoted in the lower left corner of each panel. Black and red colors represent regions on the NE and SW limbs, respectively. The x-axis is the region label as shown in Fig.~\ref{fig:NuSTARtricolor}. Regions with the same label on the NE and SW limbs, as well as the fitting results with different models for the same region, are slightly shifted on the x-axis in order to be clearly separated. $\chi^2/\rm d.o.f.$ of different models are summarized in Tables~\ref{table:NuSTARXMMpowerpara} - \ref{table:defaultpara}.}\label{fig:reducedchiAZ}
\end{center}
\end{figure*}


\begin{table*}
\begin{center}
\caption{Best-fit parameters of the srcut and sresc models} 
\footnotesize
\tabcolsep=5.0pt%
\begin{tabular}{lcccrccccccccccc}
\hline
Region & $\alpha$ (srcut) & $E_{\rm cutoff}$/keV (srcut) & $\chi^2/\rm d.o.f.$ (srcut) & $\mid$ & $\alpha$ (sresc) & $E_{\rm cutoff}$/keV (sresc) & $\chi^2/\rm d.o.f.$ (sresc) & d.o.f \\
\hline
NE\_1 & $0.458\pm0.003$ & $0.199\pm0.005$ & 1.660 & $\mid$ & $0.491\pm0.003$ & $0.854\pm0.024$ & 1.686 &  708 \\
NE\_2 & $0.474\pm0.002$ & $0.321\pm0.008$ & 1.527 & $\mid$ & $0.504\pm0.002$ & $1.350_{-0.030}^{+0.032}$ & 1.530 &  875 \\
NE\_3 & $0.479_{-0.002}^{+0.003}$ & $0.342_{-0.008}^{+0.009}$ & 1.482 & $\mid$ & $0.508\pm0.002$ & $1.444_{-0.033}^{+0.034}$ & 1.485 &  856 \\
NE\_4 & $0.468\pm0.003$ & $0.301_{-0.007}^{+0.008}$ & 1.256 & $\mid$ & $0.499\pm0.002$ & $1.287_{-0.029}^{+0.030}$ & 1.257 &  828 \\
NE\_5 & $0.471\pm0.003$ & $0.331_{-0.008}^{+0.009}$ & 1.382 & $\mid$ & $0.501\pm0.002$ & $1.410_{-0.030}^{+0.033}$ & 1.390 &  879 \\
NE\_6 & $0.470_{-0.002}^{+0.003}$ & $0.276_{-0.006}^{+0.007}$ & 1.350 & $\mid$ & $0.502\pm0.002$ & $1.186_{-0.026}^{+0.027}$ & 1.344 &  921 \\
NE\_7 & $0.465\pm0.003$ & $0.203_{-0.004}^{+0.005}$ & 1.674 & $\mid$ & $0.497\pm0.003$ & $0.861\pm0.021$ & 1.681 & 1055 \\
SW\_1 & $0.447\pm0.004$ & $0.173\pm0.005$ & 1.821 & $\mid$ & $0.478\pm0.005$ & $0.722_{-0.034}^{+0.032}$ & 1.865 &  616 \\
SW\_2 & $0.462\pm0.004$ & $0.212\pm0.007$ & 1.913 & $\mid$ & $0.498_{-0.003}^{+0.004}$ & $0.944_{-0.030}^{+0.032}$ & 1.934 &  568 \\
SW\_3 & $0.467\pm0.004$ & $0.189\pm0.008$ & 1.862 & $\mid$ & $0.501\pm0.005$ & $0.829\pm0.046$ & 1.921 &  402 \\
SW\_4 & $0.473\pm0.003$ & $0.213_{-0.006}^{+0.007}$ & 1.297 & $\mid$ & $0.504\pm0.003$ & $0.908_{-0.030}^{+0.032}$ & 1.264 &  476 \\
SW\_5 & $0.476\pm0.003$ & $0.201_{-0.006}^{+0.007}$ & 1.221 & $\mid$ & $0.507\pm0.003$ & $0.841_{-0.029}^{+0.031}$ & 1.224 &  482 \\
SW\_6 & $0.467\pm0.003$ & $0.191_{-0.005}^{+0.006}$ & 1.145 & $\mid$ & $0.498\pm0.004$ & $0.810_{-0.030}^{+0.029}$ & 1.115 &  516 \\
SW\_7 & $0.467\pm0.003$ & $0.190\pm0.005$ & 1.271 & $\mid$ & $0.496\pm0.004$ & $0.786_{-0.029}^{+0.030}$ & 1.228 &  515 \\
\hline
\end{tabular}\label{table:srcutsrescpara}
\end{center}
Similar as Table~\ref{table:Xdatabknpowerpara}, but for the parameters of the srcut and sresc models obtained by jointly fitting the \emph{NuSTAR} and \emph{XMM-Newton} spectra (panels~c,d of Figs.~\ref{fig:specXray} and \ref{fig:specotherreg}). $\alpha$ is the radio spectral index, while $E_{\rm cutoff}$ is the cutoff energy of the photon spectra.
\end{table*}

In most of the regions, the srcut, sresc, and zira2 models fit the broad-band non-thermal X-ray spectra almost equally well, and slightly better than the zira1 model (Fig.~\ref{fig:reducedchiAZ}b), which usually tends to underpredict the X-ray emission at $kT\gtrsim8\rm~keV$ (Fig.~\ref{fig:specXray}e). But in a few regions (SW 4-7), neither of the two loss-limited models (zira1 and zira2) could fit the spectra as well as the srcut and sresc models. The results are apparently inconsistent with \citet{Miceli13}, who found that the loss-limited models always give better fit to the \emph{XMM-Newton} spectra than the escape-limited models. The spectral extraction regions adopted in the present paper are much larger than those adopted in \citet{Miceli13}, because of the much lower angular resolution of \emph{NuSTAR}. Since \citet{Miceli13} focus on the brightest non-thermal filaments, we may conclude that synchrotron radiative loss is only important in X-ray band in limiting the particle acceleration at the brightest filaments.


\begin{table}
\begin{center}
\caption{Best-fit parameters of the zira1 and zira2 models} 
\footnotesize
\tabcolsep=2.8pt%
\begin{tabular}{lcccccccccccccc}
\hline
Region & $E_{\rm cut}$/keV (zira1) & $\chi^2/\rm d.o.f.$ & $\mid$ & $E_{\rm cut}$/keV (zira2) & $\chi^2/\rm d.o.f.$ & d.o.f. \\
\hline
NE\_1 & $0.192\pm0.003$ & 1.758 &  $\mid$ & $0.248\pm0.004$ & 1.693 &  708 \\
NE\_2 & $0.279_{-0.004}^{+0.005}$ & 1.647 &  $\mid$ & $0.373\pm0.007$ & 1.559 &  875 \\
NE\_3 & $0.291\pm0.005$ & 1.590 &  $\mid$ & $0.392\pm0.008$ & 1.513 &  856 \\
NE\_4 & $0.266_{-0.004}^{+0.005}$ & 1.330 &  $\mid$ & $0.355\pm0.007$ & 1.269 &  828 \\
NE\_5 & $0.287\pm0.005$ & 1.459 &  $\mid$ & $0.387_{-0.007}^{+0.008}$ & 1.400 &  879 \\
NE\_6 & $0.247\pm0.004$ & 1.434 &  $\mid$ & $0.327_{-0.005}^{+0.006}$ & 1.369 &  921 \\
NE\_7 & $0.193\pm0.002$ & 1.802 & $\mid$ & $0.249_{-0.003}^{+0.004}$ & 1.719 & 1055 \\
SW\_1 & $0.173\pm0.003$ & 1.947 &  $\mid$ & $0.222\pm0.004$ & 1.886 &  616 \\
SW\_2 & $0.197\pm0.004$ & 1.967 &  $\mid$ & $0.256_{-0.005}^{+0.006}$ & 1.931 &  568 \\
SW\_3 & $0.179\pm0.004$ & 1.970 &  $\mid$ & $0.231\pm0.006$ & 1.912 &  402 \\
SW\_4 & $0.202\pm0.004$ & 1.635 &  $\mid$ & $0.259\pm0.006$ & 1.470 &  476 \\
SW\_5 & $0.192_{-0.003}^{+0.004}$ & 1.537 &  $\mid$ & $0.245\pm0.005$ & 1.380 &  482 \\
SW\_6 & $0.189\pm0.003$ & 1.512 &  $\mid$ & $0.241_{-0.004}^{+0.005}$ & 1.327 &  516 \\
SW\_7 & $0.188\pm0.003$ & 1.658 &  $\mid$ & $0.239_{-0.004}^{+0.005}$ & 1.465 &  515 \\
\hline
\end{tabular}\label{table:zirapara}
\end{center}
Similar as Table~\ref{table:Xdatabknpowerpara}, but for the parameters of the loss-limited models (zira1 and zira2) obtained by jointly fitting the \emph{NuSTAR} and \emph{XMM-Newton} spectra (panel~e,f of Figs.~\ref{fig:specXray} and \ref{fig:specotherreg}). $E_{\rm cut}$ is the cutoff energy of the synchrotron emission spectrum.
\end{table}


\subsection{Curvature of the electron energy spectrum}\label{subsec:ElectronSpecCurv}

We further examine if the broad-band non-thermal emission from radio to hard X-ray can be described with the synchrotron emission of the same CR electron population. In order to fit the \emph{NuSTAR} and \emph{XMM-Newton} X-ray spectra together with the radio flux extracted from the flux accurate image of \citet{Dyer09}, we adopt the default built-in electron PDF of the non-thermal module of ISIS, which is responsible for the synchrotron emission from radio to X-ray \citep{Allen08}. The PDF has the form of:
\begin{equation}
\label{equi:defaultPDF}
dn/dp=A(p/p_{\rm 0})^{f(p,a)} e^{-(p-p_{\rm 0})/p_{\rm cutoff}},
\end{equation}
where the index $f(p,a)$ is:\\
$~~~~~~~~~~~~f(p,a)=-\Gamma+a\times\log(p/p_{\rm 0}),~{\rm when}~p>p_{\rm 0}$\\
$~~~~~~~~~~~~f(p,a)=-\Gamma,~{\rm when}~p\leq p_{\rm 0}$.\\
In Equ.~\ref{equi:defaultPDF}, $n$ is the electron number density, $p=\gamma mv$ ($\gamma$, $m$, and $v$ are the Lorentz factor, rest mass, and velocity of the particle), $A$ is the normalization of the PDF in unit of $p_{\rm 0}^{-1}\rm~cm^{-3}$, $p_{\rm 0}=1{\rm~GeV}~c^{-1}$ where $c$ is the light speed, $p_{\rm cutoff}$ is the cutoff momentum, $\Gamma$ and $a$ characterize the slope and curvature of the PDF, respectively.


\begin{figure*}
\begin{center}
\epsfig{figure=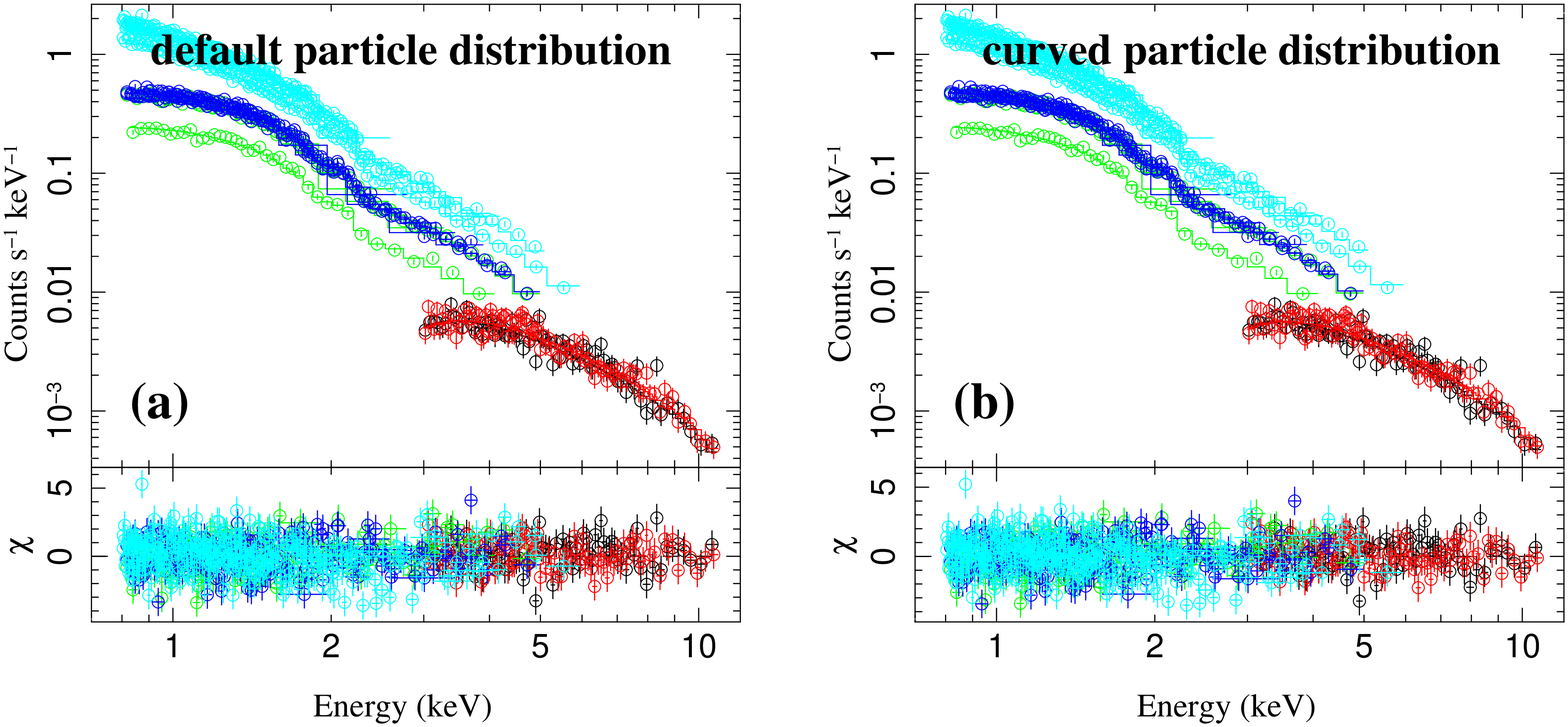,width=1.0\textwidth,angle=0, clip=}
\caption{Similar as Fig.~\ref{fig:specXray}, but in this figure the spectra from X-ray (\emph{NuSTAR} and \emph{XMM-Newton}) and radio (only one data point at 1.37~GHz, which is always well fitted) are jointly fitted with a synchrotron emission model of a power law distribution of electron energy with an exponential cutoff (``default particle distribution''). Only X-ray spectra are presented for clarity. In panel~(b), we further add a free parameter of curvature to the power law distribution of electron energy (``curved particle distribution''). Best-fit parameters are summarized in Table~\ref{table:defaultpara}.}\label{fig:specRadioXray}
\end{center}
\end{figure*}

We fit the broad-band radio-X-ray spectrum with a synchrotron emission model of a population of CR electrons with a PDF described by Equ.~\ref{equi:defaultPDF}. For a given spectrum, the total magnetic field strength ($B_{\rm tot}$) in the synchrotron emission model and the cutoff momentum in the PDF ($p_{\rm cutoff}$) are not independent (e.g., \citealt{Allen08}). We therefore need to fix one of them. There are in general two components of the magnetic field in SN1006: a large scale ordered Galactic component along the SW-NE direction or roughly parallel to the Galactic plane (so the non-thermal limbs has a polar cap geometry) and a highly disordered and turbulent component at the two non-thermal limbs which is likely responsible for the particle acceleration \citep{Reynoso13}. The strength of the disordered component in the post shock region is poorly constrained from multi-wavelength observations (e.g., \citealt{Dyer01,Allen01,Aharonian05}), but is in general $\sim100\rm~\mu G$ based on recent estimates with various methods, such as the thickness of the shock compressed filaments or the broad-band spectral analysis (e.g., \citealt{Parizot06,Morlino10,Acero10,Helder12,Berezhko12}). We therefore arbitrarily fix $B_{\rm tot}$ at $100\rm~\mu G$ in our analysis. We examine two sets of models, with or without a curvature in the PDF (i.e., $a=0$ or set free, named as ``default'' or ``curved''). An example of the best-fit spectra (only X-ray spectra are presented for clarity, the 1.37~GHz radio flux is always well fitted) are presented in Fig.~\ref{fig:specRadioXray} and the best-fit model parameters are listed in Table~\ref{table:defaultpara}. As the best-fit spectra are very similar as those fitted with an srcut model, we do not present fitted spectra of other regions.


\begin{figure}
\begin{center}
\epsfig{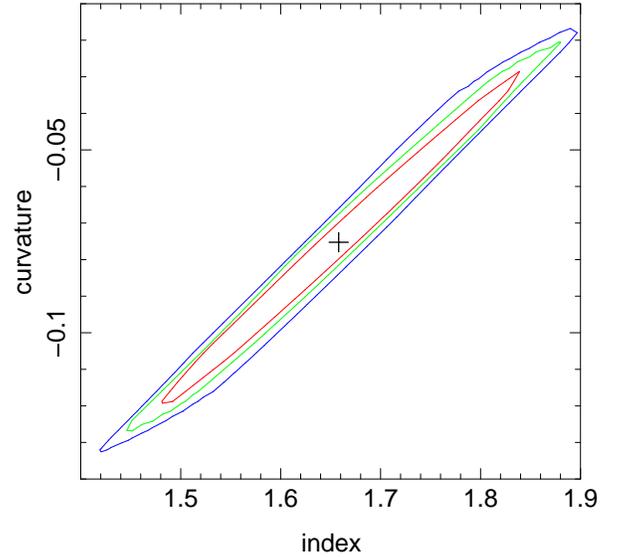}
\caption{Confidence contours of the index ($\Gamma$) and curvature ($a$) of the curved PDF (Equ.~\ref{equi:defaultPDF}) of NE region 3 (Fig.~\ref{fig:specRadioXray}b). Contours from inner to outer are at 1~$\rm \sigma$ (68.3\%; red), 1.64~$\rm \sigma$ (90\%; green), and 3~$\rm \sigma$ (99.73\%; blue) confidence levels.}\label{fig:confidenceindexcurvature}
\end{center}
\end{figure}

The radio flux and X-ray spectra can both be well fitted with such a synchrotron emission model simultaneously, indicating that they are produced by a single CR electron population. In most of the regions, the synchrotron emission of a curved or default PDF fit the spectra equally well (Fig.~\ref{fig:reducedchiAZ}c). We also found a tight correlation of the best-fit curvature and index of the PDF (Fig.~\ref{fig:confidenceindexcurvature}), and the curvature is very close to zero. Therefore, we do not find strong evidence of a curved PDF in the $2^\prime\times2^\prime$ spectral extraction regions on the non-thermal limbs of SN1006, which has been claimed for smaller regions based on the radio spectra and higher resolution \emph{Chandra} data at $\lesssim7\rm~keV$ \citep{Allen08}. However, cautions need to be made that since we only have radio flux measurement at one frequency, there is no constraint on the radio spectral slope. Furthermore, the spectral slope obtained in srcut or sresc models ($\lesssim0.5$) is always flatter than the average radio spectral index over the entire SNR ($\approx0.6$, e.g., \citealt{Dyer01,Rothenflug04,Allen08}). This may explain the difference between our results and what has been found by \citet{Allen08}. Accurate and high-resolution flux measurements at a few different frequencies in GHz band are needed to better characterize the broad-band synchrotron spectrum in SN1006.


\begin{table*}
\begin{center}
\caption{Best-fit parameters of the synchrotron model with exponentially cutoff power law (default) or curved particle distributions} 
\footnotesize
\tabcolsep=5.0pt%
\begin{tabular}{lcccccccccccccc}
\hline
Region & $\Gamma$ (default) & $E_{\rm cutoff}$/TeV (default) & $\chi^2/\rm d.o.f.$ & d.o.f & $\mid$ & $\Gamma$ (curved) & $a$ (curved) & $E_{\rm cutoff}$/TeV (curved) & $\chi^2/\rm d.o.f.$ & d.o.f \\
\hline
NE\_1 & $1.899_{-0.023}^{+0.020}$ & $5.93_{-0.10}^{+0.09}$ & 1.661 &  708 & $\mid$ & $2.095_{-0.263}^{+0.278}$ & $ 0.052_{-0.073}^{+0.075}$ & $ 5.39_{-0.62}^{+0.73}$ & 1.661 &  707 \\
NE\_2 & $1.940_{-0.021}^{+0.019}$ & $7.56\pm0.12$ & 1.523 &  875 & $\mid$ & $1.607_{-0.186}^{+0.193}$ & $-0.086_{-0.046}^{+0.049}$ & $ 9.37_{-1.10}^{+1.33}$ & 1.515 &  874 \\
NE\_3 & $1.950_{-0.021}^{+0.019}$ & $7.81\pm0.13$ & 1.479 &  856 & $\mid$ & $1.658_{-0.194}^{+0.202}$ & $-0.075_{-0.048}^{+0.052}$ & $ 9.46_{-1.17}^{+1.42}$ & 1.474 &  855 \\
NE\_4 & $1.924_{-0.022}^{+0.019}$ & $7.31\pm0.12$ & 1.254 &  828 & $\mid$ & $1.891_{-0.210}^{+0.219}$ & $-0.009_{-0.053}^{+0.057}$ & $ 7.46_{-0.84}^{+1.00}$ & 1.255 &  827 \\
NE\_5 & $1.931_{-0.022}^{+0.019}$ & $7.68\pm0.12$ & 1.380 &  879 & $\mid$ & $1.817_{-0.198}^{+0.207}$ & $-0.030_{-0.050}^{+0.054}$ & $ 8.24_{-0.94}^{+1.12}$ & 1.380 &  878 \\
NE\_6 & $1.929_{-0.022}^{+0.019}$ & $7.01_{-0.11}^{+0.10}$ & 1.350 &  921 & $\mid$ & $1.891_{-0.203}^{+0.211}$ & $-0.010_{-0.051}^{+0.055}$ & $ 7.17_{-0.76}^{+0.90}$ & 1.351 &  920 \\
NE\_7 & $1.919_{-0.022}^{+0.019}$ & $6.00\pm0.08$ & 1.671 & 1055 & $\mid$ & $1.661_{-0.213}^{+0.219}$ & $-0.067_{-0.054}^{+0.057}$ & $ 6.90_{-0.76}^{+0.90}$ & 1.669 & 1054 \\
SW\_1 & $1.878_{-0.023}^{+0.021}$ & $5.57_{-0.10}^{+0.09}$ & 1.818 &  616 & $\mid$ & $1.775_{-0.347}^{+0.366}$ & $-0.027_{-0.089}^{+0.098}$ & $ 5.87_{-0.93}^{+1.22}$ & 1.821 &  615 \\
SW\_2 & $1.907_{-0.024}^{+0.021}$ & $6.12\pm0.11$ & 1.914 &  568 & $\mid$ & $2.458_{-0.341}^{+0.366}$ & $ 0.147_{-0.091}^{+0.095}$ & $ 4.69_{-0.64}^{+0.79}$ & 1.905 &  567 \\
SW\_3 & $1.918_{-0.024}^{+0.021}$ & $5.78_{-0.12}^{+0.13}$ & 1.860 &  402 & $\mid$ & $2.260_{-0.407}^{+0.437}$ & $ 0.091_{-0.106}^{+0.117}$ & $ 4.89_{-0.81}^{+1.06}$ & 1.860 &  401 \\
SW\_4 & $1.935_{-0.022}^{+0.019}$ & $6.12\pm0.11$ & 1.291 &  476 & $\mid$ & $1.199_{-0.199}^{+0.277}$ & $-0.190_{-0.053}^{+0.071}$ & $ 9.66_{-1.63}^{+1.95}$ & 1.254 &  475 \\
SW\_5 & $1.942_{-0.022}^{+0.020}$ & $5.95\pm0.10$ & 1.213 &  482 & $\mid$ & $1.483_{-0.264}^{+0.276}$ & $-0.119_{-0.065}^{+0.072}$ & $ 7.69_{-1.12}^{+1.41}$ & 1.200 &  481 \\
SW\_6 & $1.924_{-0.021}^{+0.019}$ & $5.83_{-0.09}^{+0.10}$ & 1.136 &  516 & $\mid$ & $1.171_{-0.171}^{+0.256}$ & $-0.194_{-0.046}^{+0.065}$ & $ 9.09_{-1.36}^{+1.50}$ & 1.094 &  515 \\
SW\_7 & $1.925_{-0.021}^{+0.019}$ & $5.82\pm0.09$ & 1.263 &  515 & $\mid$ & $1.021_{-0.021}^{+0.260}$ & $-0.232_{-0.010}^{+0.063}$ & $10.13_{-1.61}^{+0.59}$ & 1.201 &  514 \\
\hline
\end{tabular}\label{table:defaultpara}
\end{center}
Parameters of the synchrotron emission model with exponentially cutoff power law (default) or curved particle distributions obtained by jointly fitting the X-ray (\emph{NuSTAR} and \emph{XMM-Newton}) and radio data (an example of the best-fit spectra are presented in Fig.~\ref{fig:specRadioXray}). $\Gamma$ is the index of the power law particle distribution function, $E_{\rm cutoff}$ is the exponential cutoff energy of the particle distribution function corresponding to the cutoff momentum $p_{\rm cutoff}$, $a$ is an additional parameter describing the curvature of the particle distribution function (Equ.~\ref{equi:defaultPDF}). The magnetic field has been fixed at 100$\rm~\mu G$ because it is directly linked to $E_{\rm cutoff}$.
\end{table*}

\section{Discussions}\label{sec:Discussions}


\subsection{Shape of the broad-band synchrotron emission and CR energy distribution}\label{subsec:syncshape}

Standard diffusive shock acceleration (DSA) theory predicts the index of the energy spectrum of the accelerated particles to be $\Gamma=2$ with a shock compression ratio of $r=4$ (e.g., \citealt{Blandford87}). In recent years, it has been noticed that CRs could not be simply regarded as test particles, but must carry non-negligible amount of energy and pressure and actively participate the shock dynamics (e.g., \citealt{Blandford87,Caprioli12}). Such non-linear theories of DSA predict back-reaction of the accelerated particles which form a precursor in the upstream and slow down the fluid. The net effect is that higher energy particles ``feel'' a larger compression ratio and are accelerated more efficiently. The resultant PDF is thus concave, with an index $\Gamma>2$ at low energy and $\Gamma<2$ at high energy.

Such a concavity in the PDF has not yet been clearly evidenced at high confidence level. \citet{Allen08} made use of the radio and \emph{Chandra} X-ray observations of SN1006 to find that the synchrotron spectrum extracted from much smaller regions than what have been adopted in the present paper seem to flatten with increasing energy. For comparison, we have obtained an index of the ``default'' PDF for the NE limb typically in the range of 1.9-1.95 with a mean value of $\Gamma=1.927_{-0.008}^{+0.007}$ ($\Gamma=1.919_{-0.009}^{+0.008}$ for the SW limb; 1-$\sigma$ confidence; Table~\ref{table:defaultpara}), which is even lower than the value of $\Gamma=2.073_{-0.020}^{+0.021}$ (90\% confidence) obtained from the \emph{Chandra} and radio data also for the NE limb by \citet{Allen08}. This is probably because we have included the higher energy \emph{NuSTAR} data in the analysis which makes the CR energy spectra even flatter. $\Gamma$ in all of the regions based on the ``default'' PDF is also lower than the expected value (2.0) from the standard DSA. These flat synchrotron spectra are apparently consistent with the above scenario of non-linear DSA, but we do not find a significant concavity of the PDF as claimed by \citet{Allen08}. Oppositely, the synchrotron emission spectrum shows significant curvature as indicated by the mathematical models in \S\ref{subsec:XrayCurv}. Nevertheless, the scatter of the curvature $a$ of the PDF in Equ.~\ref{equi:defaultPDF} is too large and on average either negative, or more conservatively to say, not inconsistent with no curvature (the mean value for the NE/SW limb is $a=-0.032_{-0.020}^{+0.022}$/$a=-0.075_{-0.027}^{+0.032}$ at 1-$\sigma$ confidence compared to \citealt{Allen08}'s value of $0.054\pm0.006$ at 90\% confidence for the NE limb; Fig.~\ref{fig:syncAZ}c, Table~\ref{table:defaultpara}). Furthermore, we also find a strong dependence of $a$ on the index $\Gamma$ of the PDF (Fig.~\ref{fig:confidenceindexcurvature}), suggesting that we cannot simultaneously well constrain these two parameters with the current radio/X-ray spectra. Since the ``default'' PDF can fit the data equally well as the ``curved'' PDF, we conclude that the apparently curved synchrotron emission is a natural result of the synchrotron emission of an exponentially cutoff PDF with no significant concavity or curvature.

\citet{Miceli13} analyzed the same \emph{XMM-Newton} dataset as adopted in the present paper, but extracted spectra from some smaller regions covering the brightest filaments in the non-thermal limbs. They found that the loss-limited models (zira1 and zira2) in general give better fit to the \emph{XMM-Newton} spectra than the srcut model. This is a good evidence for efficient synchrotron radiative loss which may change the shape of the non-thermal spectrum, at least in the brightest filaments. However, we do not obtain the same result, as the srcut (or sresc) model always gives a slightly better fit to the X-ray (\emph{XMM-Newton} and \emph{NuSTAR}) spectra than the loss-limited models (Fig.~\ref{fig:reducedchiAZ}b). This is probably because we have extracted spectra from larger regions including those with less efficient synchrotron loss. Alternatively, a stronger non-linear effect at higher energy (\emph{NuSTAR} band) may flatten the synchrotron spectrum, and make the loss-limited models underpredict the synchrotron emission at high energy (the e and f panels of Figs.~\ref{fig:specXray}, \ref{fig:specotherreg}).

The shape of the PDF can be affected by many processes. As discussed above, the non-linear DSA tends to produce a concave PDF with $a>0$, inconsistent with our observed value of $a\lesssim0$. However, other processes may mitigate or even remove the curvature. These processes include, but not limited to: efficient synchrotron radiative cooling at high energy (e.g., \citealt{Zirakashvili07}), density inhomogeneities in the downstream (e.g., \citealt{Fraschetti18}), turbulent heating in the CR precursor (e.g., \citealt{Berezhko99}), and enhanced magnetic field amplification induced by CR streaming in the upstream and the corresponding reduced compressibility of the plasma (e.g., \citealt{Caprioli09,Caprioli11}).


\subsection{Spatial variation of non-thermal emission parameters}\label{subsec:spatialvariation}

\begin{figure*}
\begin{center}
\epsfig{figure=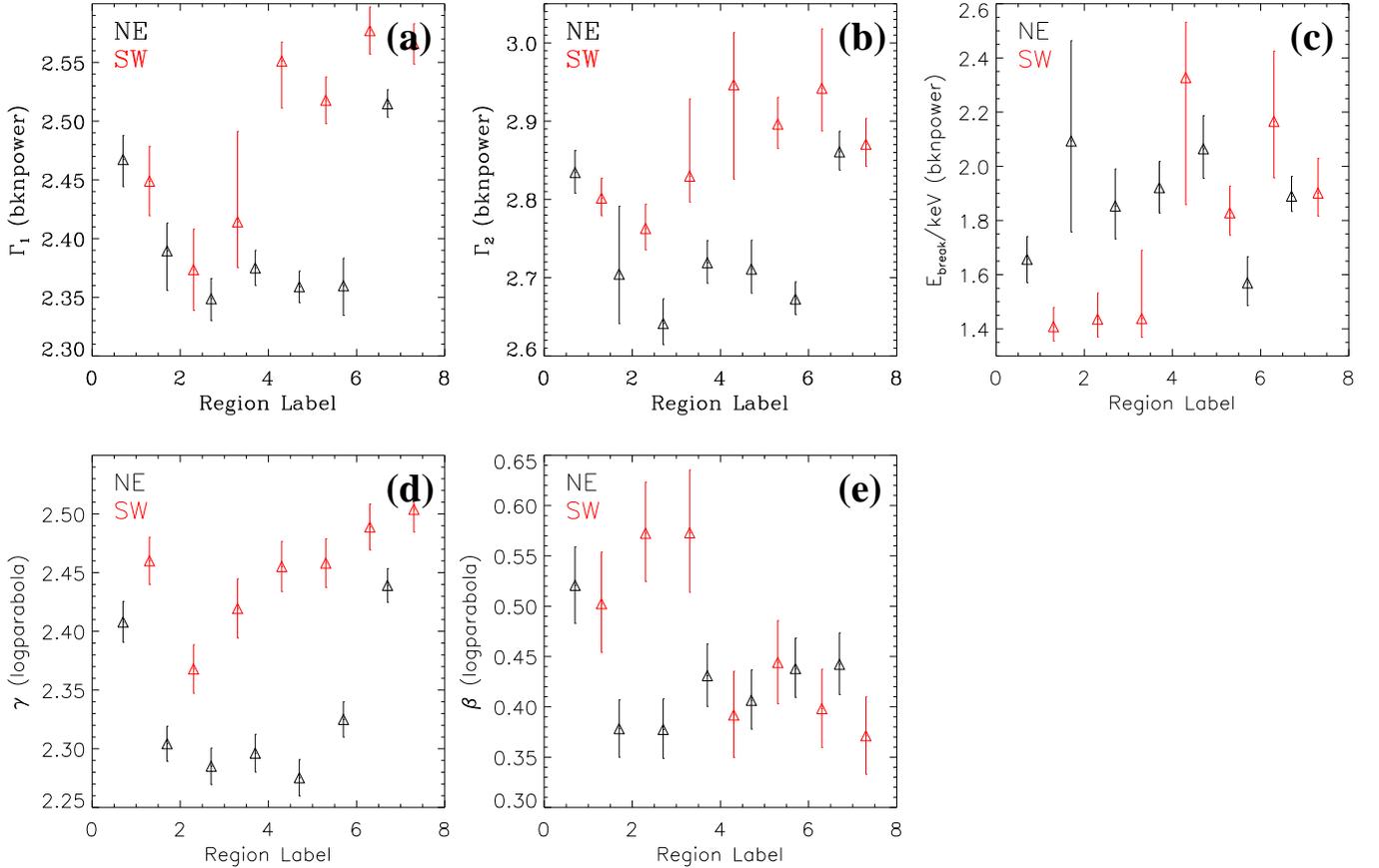,width=1.0\textwidth,angle=0, clip=}
\vspace{-0.3in}
\caption{Azimuthal variation of the parameters of the mathematical models fitted to only the X-ray (\emph{NuSTAR} and \emph{XMM-Newton}) spectra. The top row shows the parameters of the broken power law, while the bottom row shows the parameters of the log-parabolic model. The best-fit spectra are presented in panels (a-b) of Figs.~\ref{fig:specXray} and \ref{fig:specotherreg}, while the model parameters are listed in Tables~\ref{table:Xdatabknpowerpara}, \ref{table:Xdatalogparabolapara}.}\label{fig:mathAZ}
\end{center}
\end{figure*}

\begin{figure*}
\begin{center}
\epsfig{figure=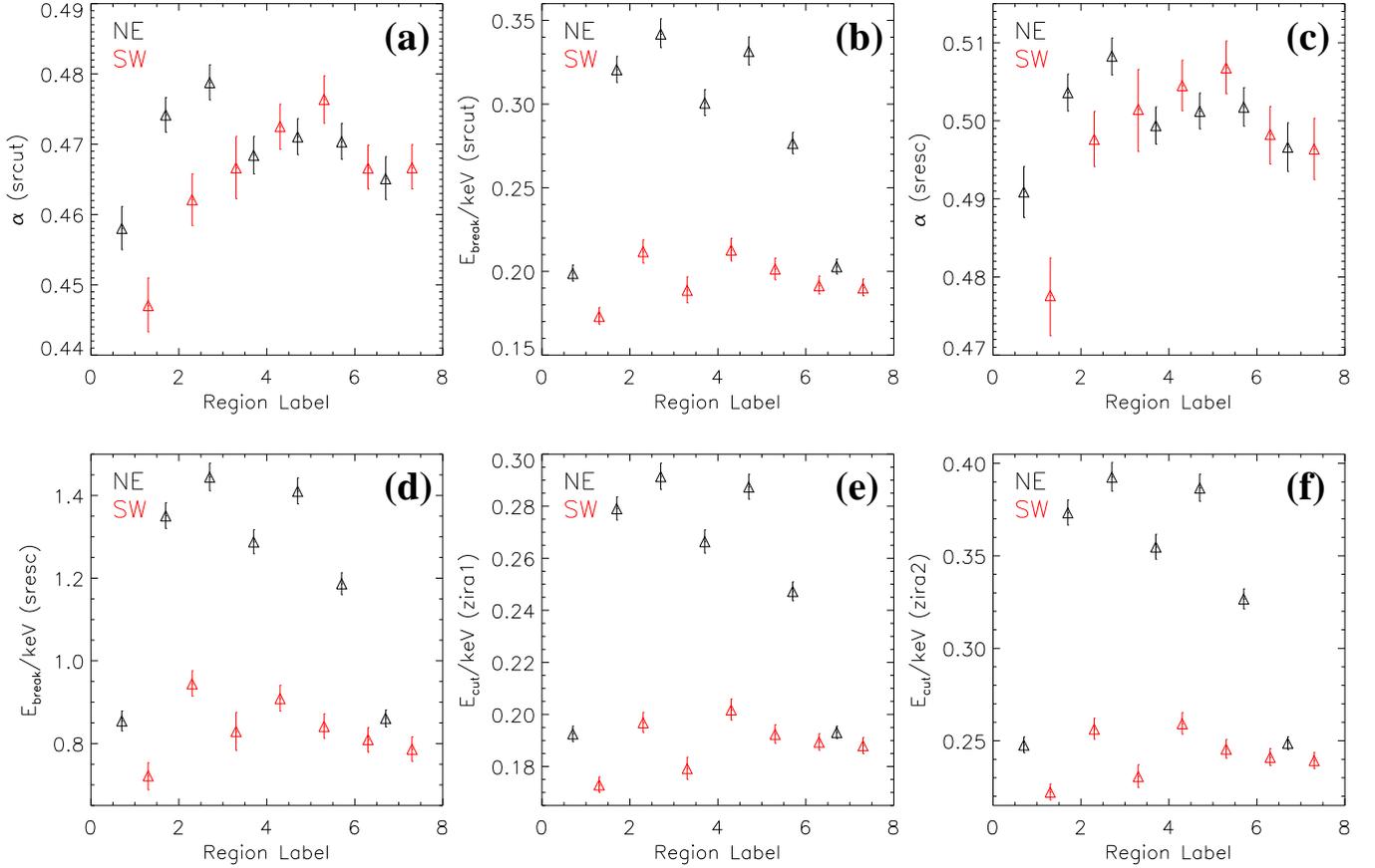,width=1.0\textwidth,angle=0, clip=}
\vspace{-0.3in}
\caption{Azimuthal viariation of the parameters of the srcut, sresc, zira1, and zira2 models fitted to only the X-ray spectra. The best-fit spectra are presented in panels~(c-f) of Figs.~\ref{fig:specXray} and \ref{fig:specotherreg}, while the model parameters are listed in Tables~\ref{table:srcutsrescpara} and \ref{table:zirapara}.}\label{fig:srcutsrsecziraAZ}
\end{center}
\end{figure*}

\begin{figure*}
\begin{center}
\epsfig{figure=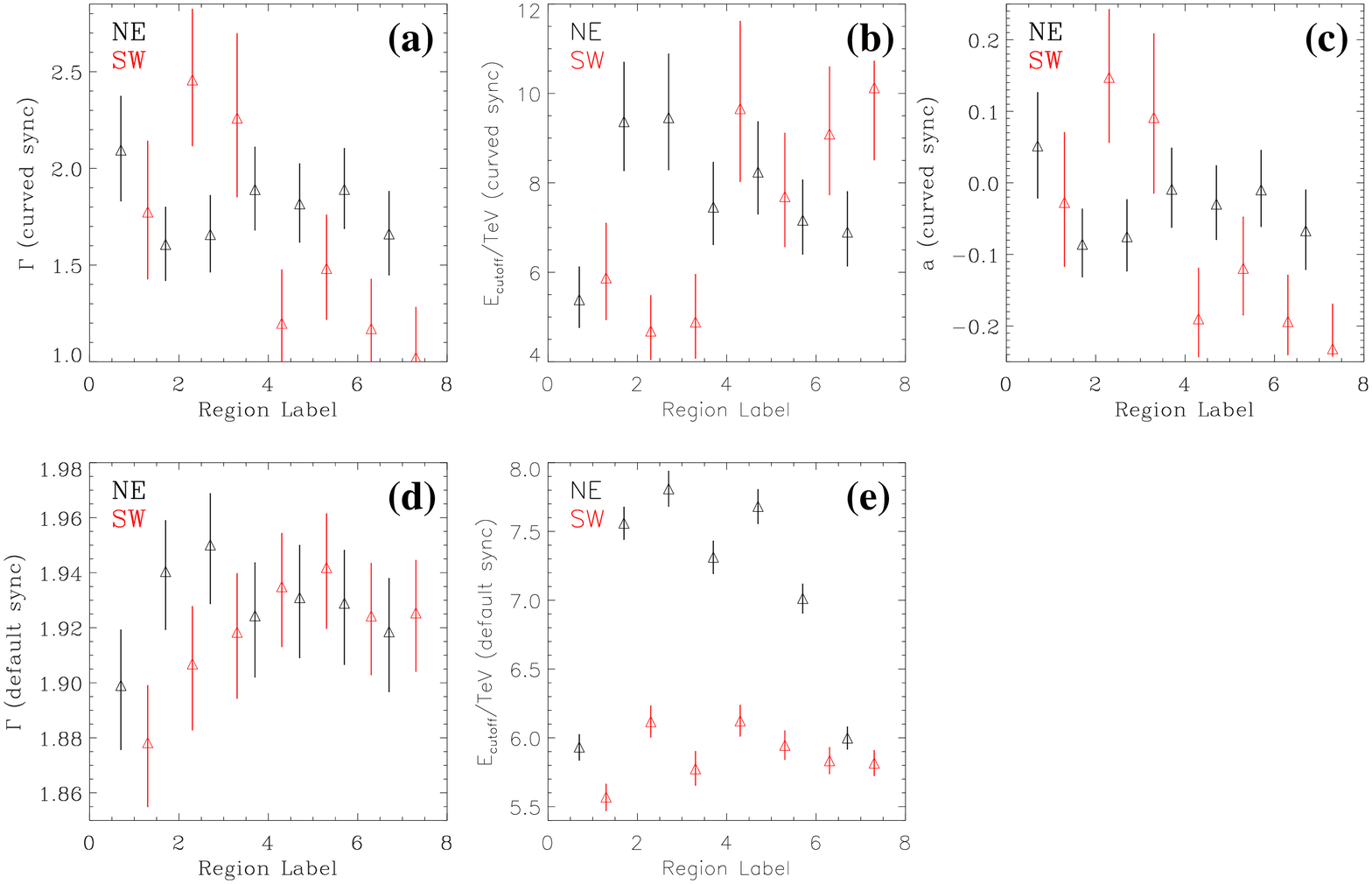,width=1.0\textwidth,angle=0, clip=}
\vspace{-0.3in}
\caption{Azimuthal viariation of the parameters of the synchrotron emission model with default (``default sync'') or curved (``curv sync'') particle energy distributions fitted to both the X-ray and radio spectra. The best-fit spectra are presented in Fig.~\ref{fig:specRadioXray}, while the model parameters are listed in Table~\ref{table:defaultpara}.}\label{fig:syncAZ}
\end{center}
\end{figure*}

\begin{figure}
\begin{center}
\epsfig{figure=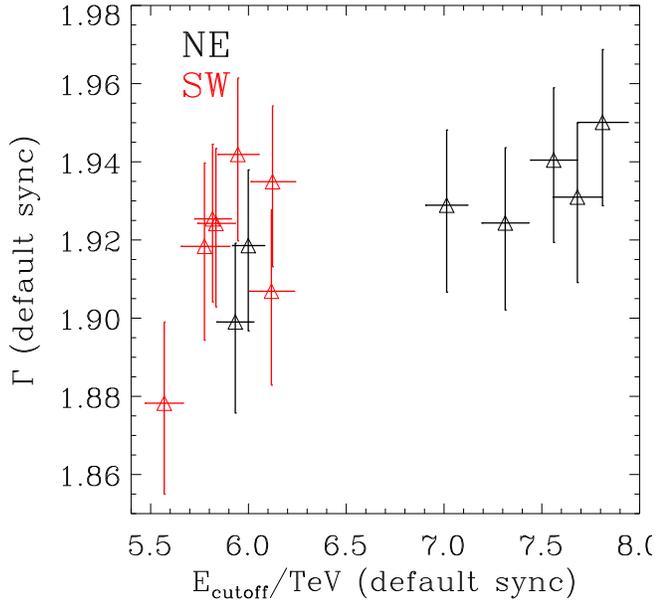,width=0.48\textwidth,angle=0, clip=}
\vspace{-0.3in}
\caption{Index ($\Gamma$) and cutoff energy ($E_{\rm cutoff}$) of the ``default'' PDF.}\label{fig:EcutoffGamma}
\end{center}
\end{figure}

We present the azimuthal variation of various spectral model parameters along the non-thermal limbs in Figs.~\ref{fig:mathAZ}, \ref{fig:srcutsrsecziraAZ}, \ref{fig:syncAZ}. The mathematical models indicate that the non-thermal spectrum is in general flatter (smaller photon index $\Gamma$ for bknpower and $\gamma$ for logparabola) in the center of the NE limb (region 4) and becomes steeper at larger azimuthal angles (from 4 to 1 or 4 to 7; Fig.~\ref{fig:mathAZ}). The spectra also tend to be flatter in the NE limb than in the SW limb. On the other hand, the other parameters of the mathematical models, such as the break energy of bknpower and curvature $\beta$ of logparabola, do not show significant azimuthal variations. The steepening of the non-thermal spectra from the center of the limb to large azimuthal angles, however, could be explained by a higher cutoff energy instead of a smaller photon index of any physical models in the center of the limb (the photon index $\alpha$ is indeed larger in the center, Fig.~\ref{fig:srcutsrsecziraAZ}; also revealed in previous works: e.g., \citealt{Rothenflug04,Miceli09,Miceli14a,Miceli16}). This is also consistent with the shape of the PDF, i.e., both higher cutoff energy $E_{\rm cutoff}$ and steeper slope $\Gamma$ in the center of the limb where the particle acceleration seems the most efficient (Fig.~\ref{fig:syncAZ}d,e). $E_{\rm cutoff}$ and $\Gamma$ are in general positively correlated with each other (Fig.~\ref{fig:EcutoffGamma}). Consistent with discussed in \S\ref{subsec:syncshape}, this correlation suggests that when the turbulent magnetic field is stronger \citep{Reynoso13} and the CR acceleration is the most efficient, the non-linear effect tends to be the most significant, making the particles accelerated to higher energy. In the mean time, the PDF becomes steeper which may be a result of many processes as listed at the end of \S\ref{subsec:syncshape}.

SN1006 is close enough that proper motion measurements over a timeline of $\sim10\rm~yrs$ has led to an accurate measurement of the expansion velocity of a few bright filaments (e.g., \citealt{Katsuda09,Winkler14}). In this paper, based on the relatively low-resolution \emph{NuSTAR} data, we are discussing the synchrotron emission in much larger regions than the width of the filaments with expansion velocity measured, so we do not have a quantitative comparison between the synchrotron emission parameters and the expansion velocity. Nevertheless, the overall trend of the azimuthal variations of the cutoff energy of the PDF and the expansion velocity seems quite different from what has been found in the Tycho's SNR, where they show a clear positive correlation which indicates that the CR acceleration is age-limited instead of loss-limited in this much younger and also Type~Ia SNR \citep{Lopez15}. Oppositely, the cutoff energy of the PDF in SN1006, at least in the NE limb, seems to be anti-correlated with the expansion velocity \citep{Winkler14}. Further investigations of the spatial variations and the relation between the cutoff energy and the shock velocity require high resolution maps of the synchrotron emission parameters which could resolve individual filaments. Right now based on the \emph{NuSTAR} data, we speculate that the efficient CR acceleration in the non-thermal limbs may have decelerated the shock significantly, producing the apparent anti-correlation between the cutoff energy and the expansion velocity. This idea is supported by the fact that the highest expansion velocity is observed along the southeast periphery of the SNR, where little non-thermal emission has been detected \citep{Winkler14}.


We do not find any significant correlations between the radio (or hard X-ray) flux and the spectral fitting parameters (such as the cutoff frequency of srcut) as revealed by previous works (e.g., \citealt{Katsuda10,Li15}). This is probably because the low angular resolution of \emph{NuSTAR} has smoothed the small scale variation (as small as the size of the filaments) of synchrotron emission parameters.


The cutoff energy in both the emission spectrum and the PDF are significantly higher in the NE limb than in the SW limb (Figs.~\ref{fig:srcutsrsecziraAZ}b,d,e,f, \ref{fig:syncAZ}e), apparently indicating that the particle acceleration is more efficient in the NE limb. Furthermore, we have also noticed that the slope of both the emission spectrum and the default PDF in SW regions 1-3 is systematically flatter than those in regions with similar angular distances from the center of the limb (Figs.~\ref{fig:srcutsrsecziraAZ}a,c, \ref{fig:syncAZ}d). The $\chi^2/\rm d.o.f.$ of any models of these three regions is also systematically higher than other regions (Fig.~\ref{fig:reducedchiAZ}), with nearly all of the models underpredict the X-ray emission at $\gtrsim8\rm~keV$ (Fig.~\ref{fig:specotherreg}). The SNR shocks in these regions have been reported to interact with an \ion{H}{1} cloud, producing slighly increased absorption column density and decreased cutoff frequency \citep{Miceli14a}. It is thus likely that the encounter of a high density cloud in the northern part of the SW limb has significantly slowed down the shock (consistent with proper motion measurements of the non-thermal filaments, e.g., \citealt{Winkler14}) and made the particle acceleration less efficient. In the mean time, the interaction with the cloud has created some non-linear effects, making the PDF flatter and the radiative cooling less efficient.


\subsection{Compare with $\gamma$-ray observations}\label{subsec:CRenergybudget}

Type~Ia SNRs are typically not strong $\gamma$-ray emitters because of the relatively low ambient density. SN1006 is one of the few Type~Ia SNRs well resolved in both TeV by ground-based Cherenkov Telescopes (e.g., \citealt{Acero10}) and in GeV by \emph{Fermi} LAT (e.g., \citealt{Xing16,Condon17}).

Our \emph{NuSTAR} observations for the first time confirm that the energy distribution of the CR electrons responsible for the synchrotron X-ray emission beyond 10~keV does not show a significant concavity or curvature, and could be described with a single power law with an exponential cutoff. The emission spectrum, however, is significantly steeper than those in GeV ($\Gamma\approx1.79$, \citealt{Condon17}) and TeV ($\Gamma\approx2.3$, \citealt{Acero10}) bands (Fig.~\ref{fig:PhoIndexXMMNuSTAR}). Therefore, although a pure leptonic scenario based on inverse Compton (IC) emission of the same electron population can well describe the GeV spectrum, it tends to under-predict the emission at TeV \citep{Acero10}. On the other hand, a pure hadronic model requires a flatter proton spectrum than the electron spectrum in order to match the flatter TeV emission spectrum. The required magnetic field is also consistent with our assumptions of $B_{\rm tot}\sim100\rm~\mu G$. However, such a model requires a very large fraction ($\sim20\%$) of the SN energy goes into CRs, which is close to the upper limit, if not unexpected. A mixed model, with the leptonic component dominating the low-energy $\gamma$-ray range [typically $\lesssim(0.1-1)\rm~TeV$], and the hadronic component (via the decay of $\pi^0$ meson) dominating the higher energy range, is probably more reasonable to describe the broad-band non-thermal spectrum \citep{Acero10}. Such a model requires an even higher cutoff energy of the CR protons than the pure hadronic model, which is $\sim100\rm~TeV$, or more than 10~times of the cutoff energy of CR electrons as revealed by our X-ray observations (Table~\ref{table:defaultpara}). The maximum energy of CR electrons is thus likely largely limited by synchrotron radiative loss which does not affect the CR hadrons.

Although the TeV morphology of SN1006 seems roughly symmetric \citep{Acero10}, the GeV emission is much stronger in the NE limb than in the SW limb, with the latter not even firmly detected \citep{Condon17}. Although the CR electron populations responsible for the GeV (via IC loss) and hard X-ray emissions (via synchrotron loss) may not be exactly the same, the significant asymmetry in GeV flux is in general consistent with our X-ray measurements, which indicates much lower cutoff energy in the SW limb probably due to the interaction with the denser ISM in this direction (Fig.~\ref{fig:syncAZ}e). The significant difference in the GeV flux and the consistency of the TeV spectra in the NE and SW limbs are difficult to be explained with a single leptonic or hadronic scenario.


\section{Summary and Conclusions}\label{section:Summary}

We present deep \emph{NuSTAR} observations of the NE and SW non-thermal limbs of the Galactic SNR SN1006. We discover three sources with X-ray emission detected at $\gtrsim50\rm~keV$. Two of them have significant multi-wavelength counterparts and are identified as background AGN, while one is not identified with online catalogues and is likely a Galactic source. We also extract the \emph{NuSTAR} spectra from a few regions along the non-thermal limbs and jointly analyze them with the \emph{XMM-Newton} spectra at $kT\gtrsim0.8\rm~keV$ and the radio data corrected for the missing flux. Below we summarize our major results and conclusions on the non-thermal emission from the two limbs:

$\bullet$ The X-ray spectral slope is clearly steeper in the \emph{NuSTAR} band than in the \emph{XMM-Newton} band. The broad-band X-ray spectra cannot be characterized with a power law with just a single break. Instead, they can be characterized with a curved power law model (``logparabola''). When fitted with a synchrotron emission model, however, the shape of the broad-band X-ray spectra is consistent with the synchrotron emission from a single population of CR electrons with a power law energy distribution and an exponential cutoff. The power law index of the PDF is typically $1.88-1.95$ for both limbs, but the cutoff energy is significantly higher in the NE limb ($\sim7.5\rm~TeV$ v.s. $\sim6.0\rm~TeV$), assuming the same magnetic field of $\sim100\rm~\mu G$. The data presented in this paper, with the radio flux only measured at one frequency, do not support a significant variation of the slope of the PDF at different energies on a scale of $\sim2^\prime$, although a concavity of the PDF is previously claimed with the radio spectra and the higher-resolution \emph{Chandra} data. 

$\bullet$ The loss-limited models do not provide a better fit to the non-thermal X-ray spectra than the escape-limited models, as suggested by \citet{Miceli13} in the study of a few brightest non-thermal filaments. In some regions in the SW limb, the loss-limited models always under predict the X-ray emissions at $\gtrsim8\rm~keV$. Because of the lower angular resolution of \emph{NuSTAR}, we have to extract spectra from larger regions covering both the highly structured filaments and the diffuse non-thermal emission from the limbs. Therefore, we conclude that synchrotron radiative losses are only important in the X-ray band in limiting the particle acceleration at the brightest filaments, which apparently have the most compressed and amplified magnetic field so the highest cooling rate.

$\bullet$ Some of the synchrotron emission parameters show significant spatial variations. The broad-band X-ray spectrum is in general flatter in the center of the NE and SW limbs and becomes steeper at larger azimuthal angles. The flatter emission spectrum in the center of the limb, however, is caused by a significantly higher cutoff energy of the CR electron PDF, instead of a flatter slope of the PDF. The cutoff energy also seems to be anti-correlated with the expansion velocity measured from the proper motion of some filaments. This trend is opposite to what has been found in the Tycho's SNR. In addition to the gradual azimuthal variation, we also find that the slopes of both the emission spectrum and the PDF are significantly smaller in three regions in the SW limb where the shock encounters a higher density ambient medium. Nearly all of the models underpredict the X-ray emission at $\gtrsim8\rm~keV$ in these three regions, indicating a more complicated PDF of the accelerated CRs. Furthermore, the NE limb shows significantly higher cutoff energies in the PDF than the SW limb, indicating more efficient particle acceleration, which is consistent with the much brighter GeV $\gamma$-ray emission on this side. On the other hand, the TeV emission, which is largely contributed by the hadronic emission, is much more symmetric on the two non-thermal limbs. The morphology of the broad-band non-thermal emission indicates that the asymmetry in the ambient medium and magnetic fields may have largely modified the leptonic CR emissions.

\bigskip
\noindent\textbf{\uppercase{acknowledgements}}
\smallskip\\
\noindent The authors would like to acknowledge Kristy Dyer and Stephen Reynolds for providing their radio images of SN1006, Daniel Wik for his very helpful suggestions on \emph{NuSTAR} background analysis, John Houck, Shuinai Zhang, Glenn E. Allen, and Michael Nowak for their helps and discussions on the non-thermal emission models and the usage of ISIS, as well as Federico Fraschetti for scientific discussions. JTL acknowledges the financial support from NASA through the grants NNX15AV24G, NNX15AM93G, SOF05-0020, 80NSSC18K0536, and NAS8-03060. The research leading to these results has received funding from the European Union's Horizon 2020 Programme under AHEAD project (grant agreement n. 654215).

\begin{appendices}

\section{Fitted spectra of all regions}\label{appendsec:SpecAllReg}

We herein present the spectra extracted from all the source regions shown in Fig.~\ref{fig:NuSTARtricolor}. The spectra are jointly fitted with the radio flux with srcut, sresc, zira1, zira2, default sync, and curv sync models.

\begin{figure*}
\begin{center}
\epsfig{figure=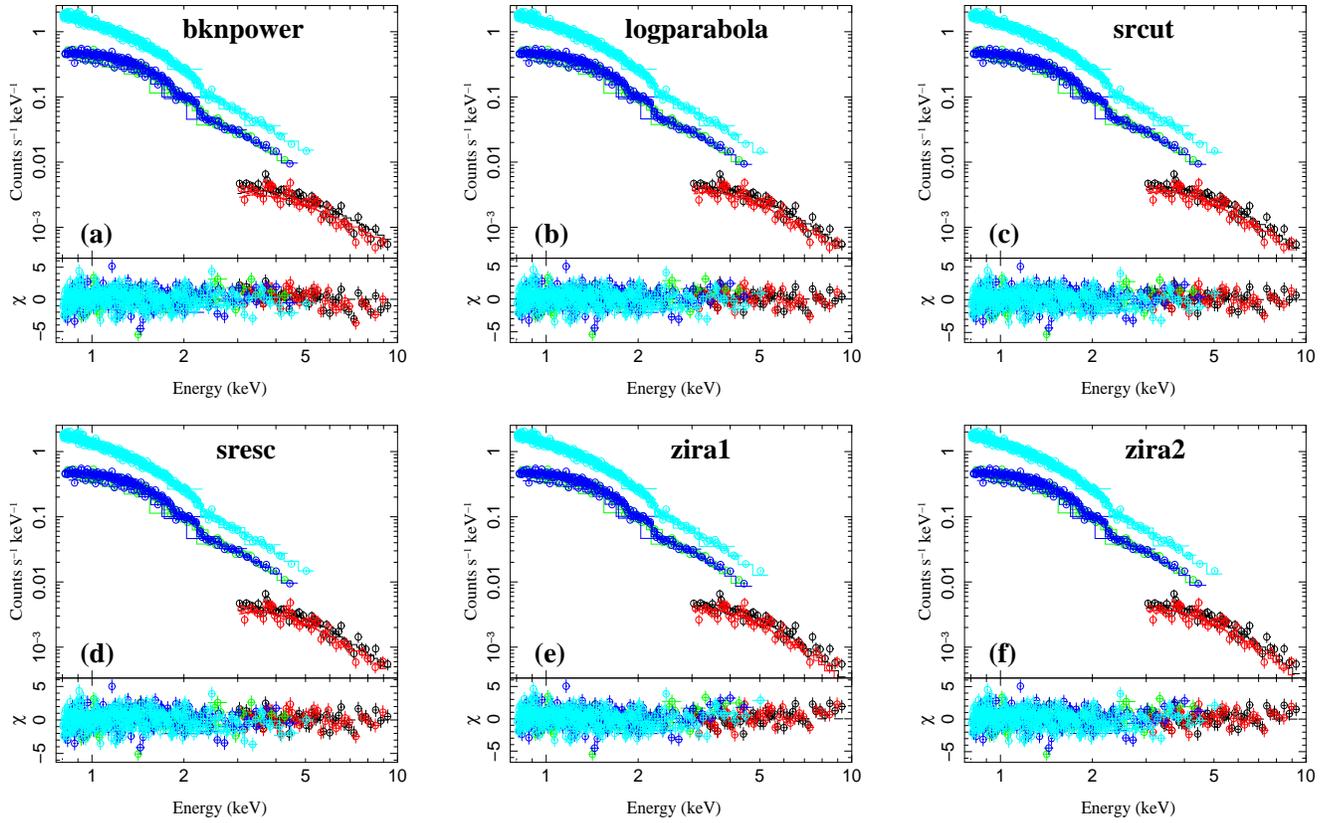,width=1.0\textwidth,angle=0, clip=}
\vspace{-0.3in}
\caption{Similar as Fig.~\ref{fig:specXray}, but for region ``1'' in the NE limb. Best-fit parameters of different models are summarized in Tables~\ref{table:NuSTARXMMpowerpara} - \ref{table:zirapara}.}\label{fig:specotherreg}
\end{center}
\end{figure*}
\addtocounter{figure}{-1}
\begin{figure*}
\begin{center}
\epsfig{figure=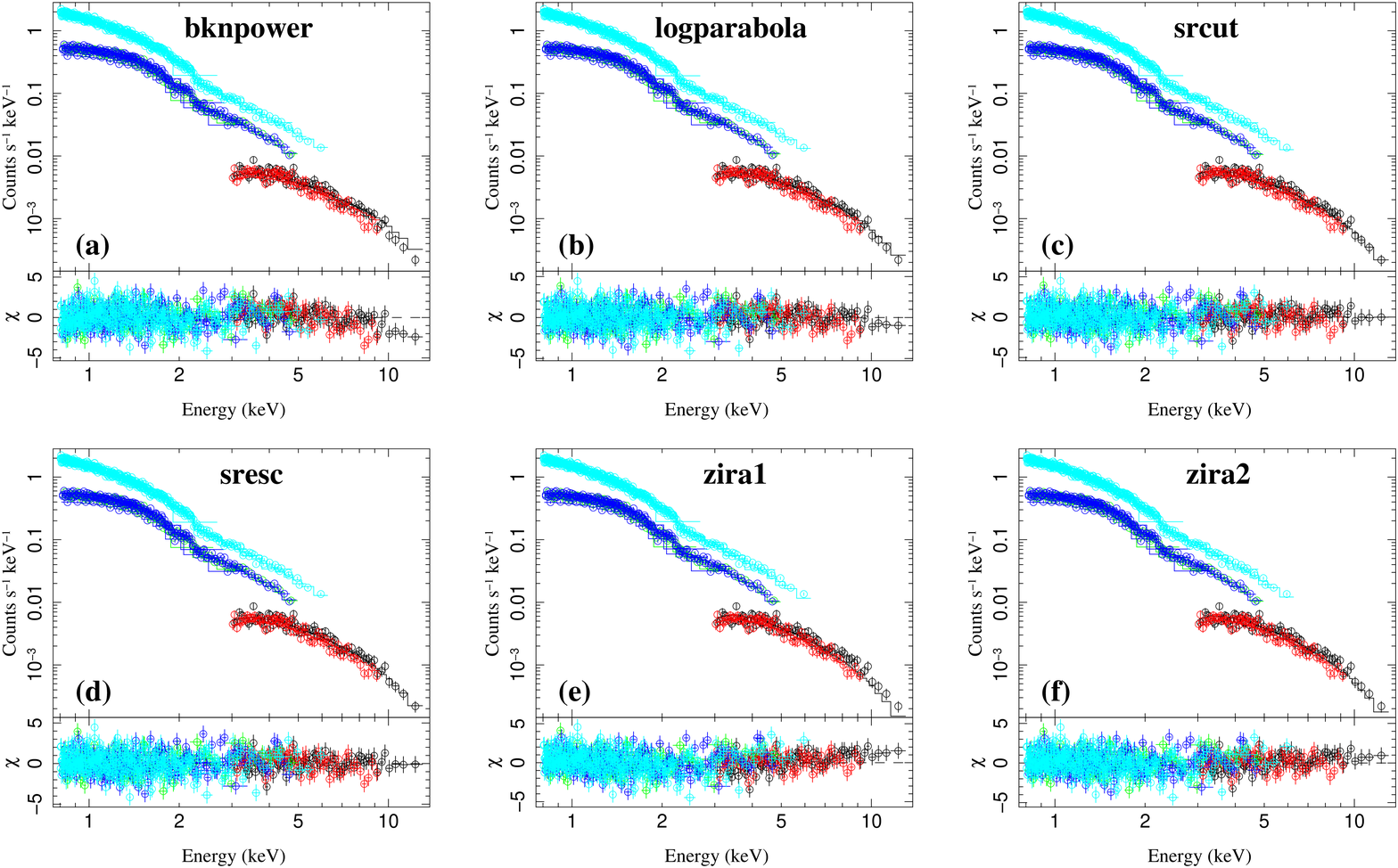,width=1.0\textwidth,angle=0, clip=}
\vspace{-0.3in}
\caption{continued, for region ``2'' in the NE limb.}
\end{center}
\end{figure*}
\addtocounter{figure}{-1}
\begin{figure*}
\begin{center}
\epsfig{figure=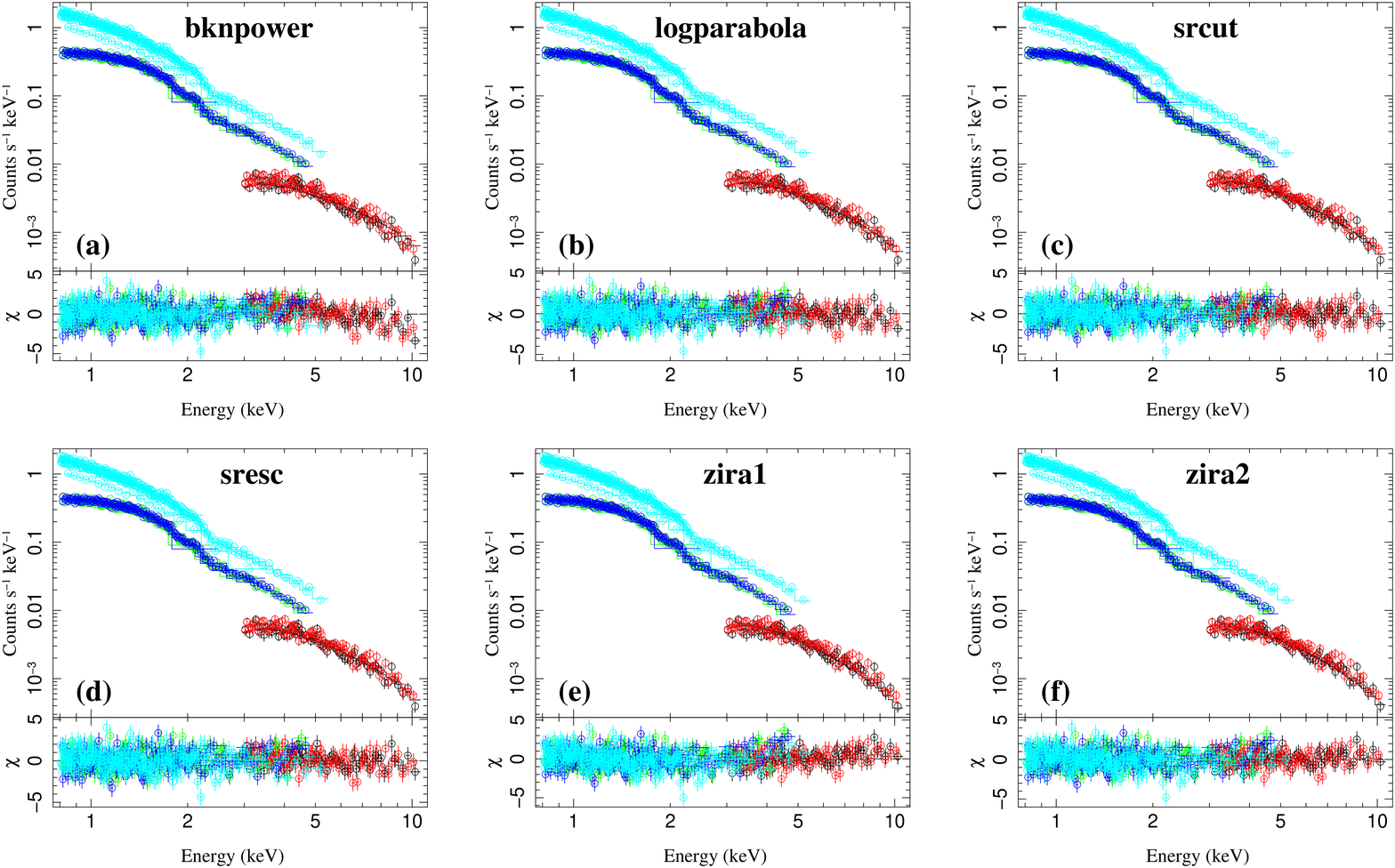,width=1.0\textwidth,angle=0, clip=}
\vspace{-0.3in}
\caption{continued, for region ``4'' in the NE limb.}
\end{center}
\end{figure*}
\addtocounter{figure}{-1}
\begin{figure*}
\begin{center}
\epsfig{figure=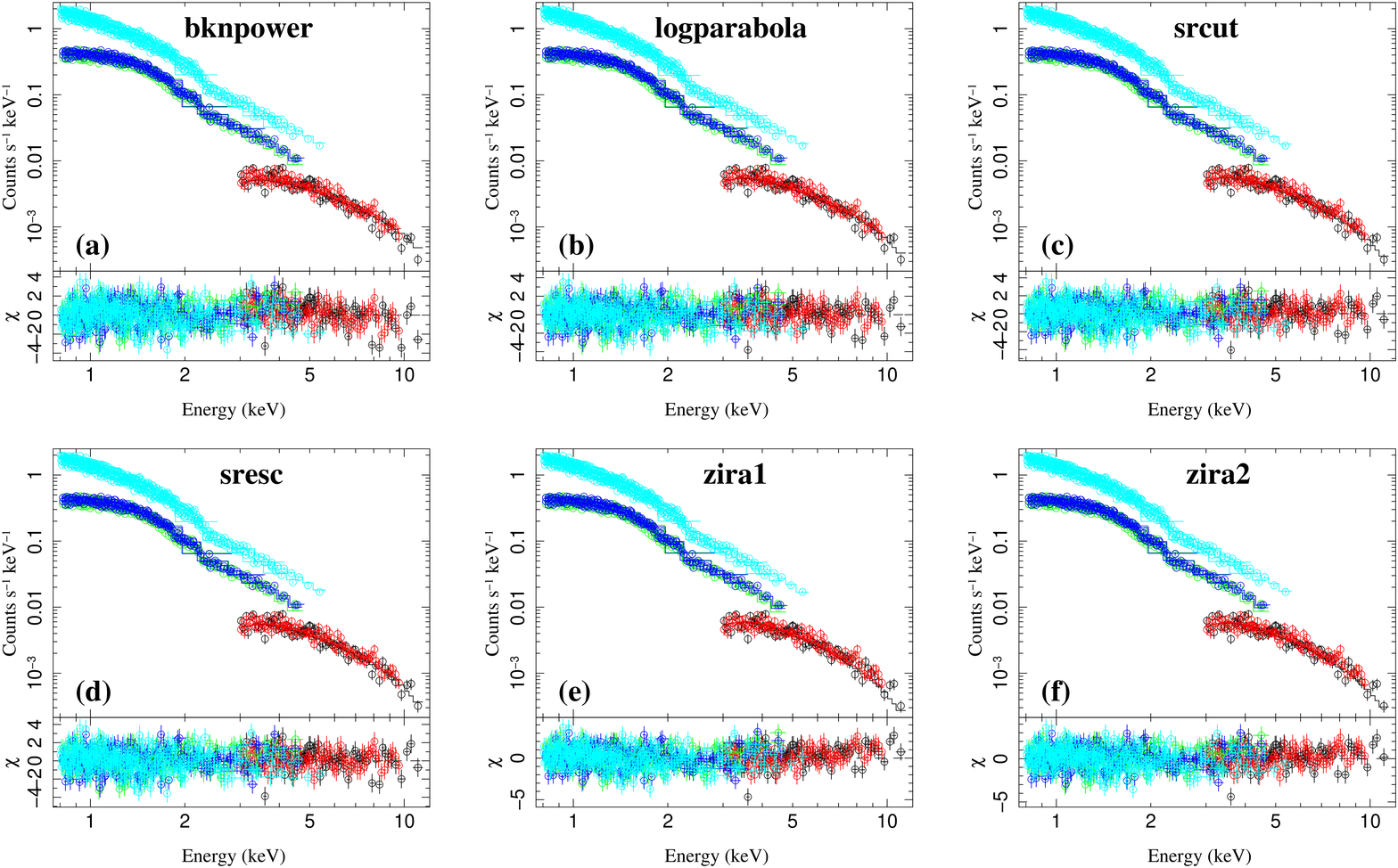,width=1.0\textwidth,angle=0, clip=}
\vspace{-0.3in}
\caption{continued, for region ``5' in the NE limb.}
\end{center}
\end{figure*}
\addtocounter{figure}{-1}
\begin{figure*}
\begin{center}
\epsfig{figure=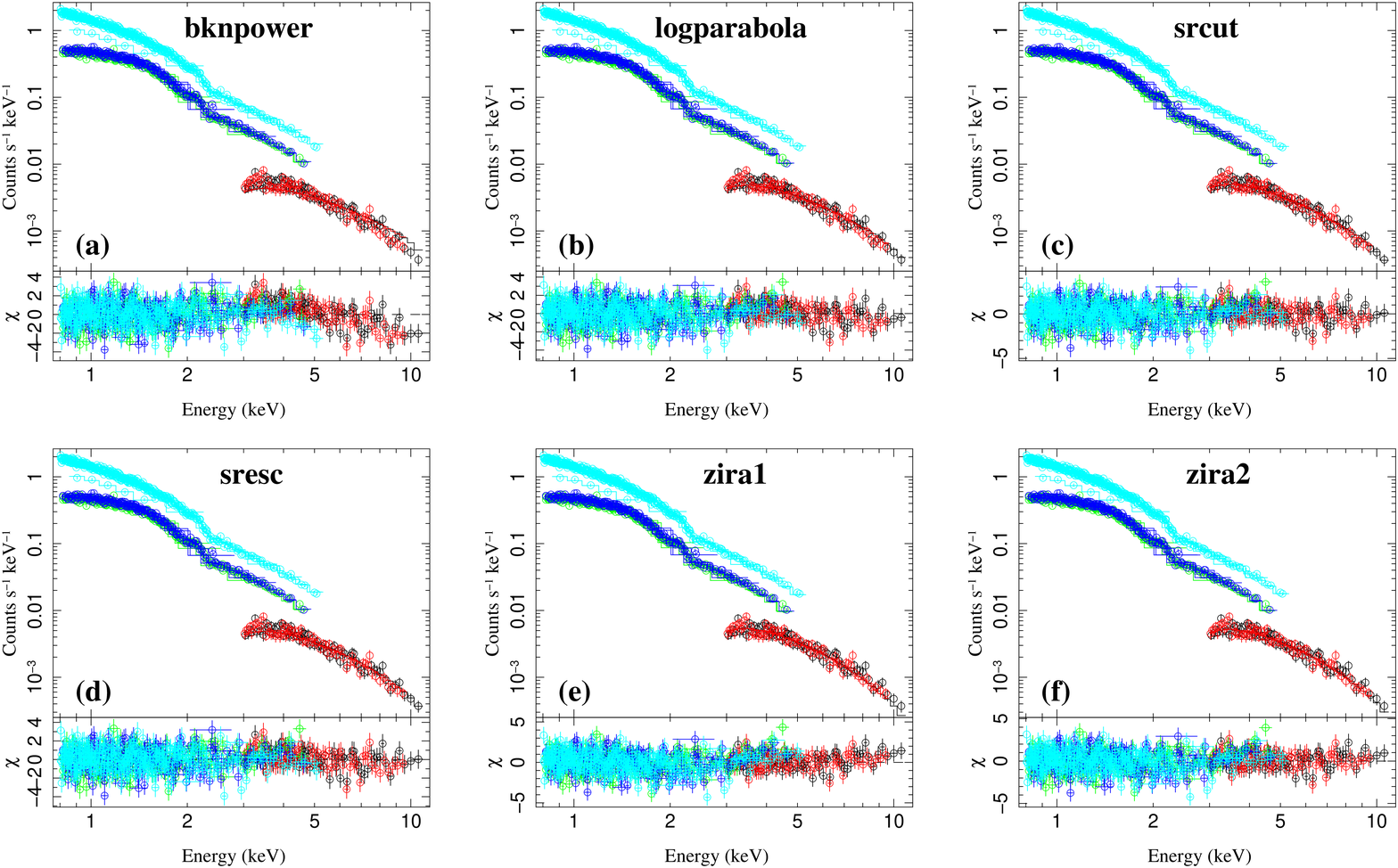,width=1.0\textwidth,angle=0, clip=}
\vspace{-0.3in}
\caption{continued, for region ``6'' in the NE limb.}
\end{center}
\end{figure*}
\addtocounter{figure}{-1}
\begin{figure*}
\begin{center}
\epsfig{figure=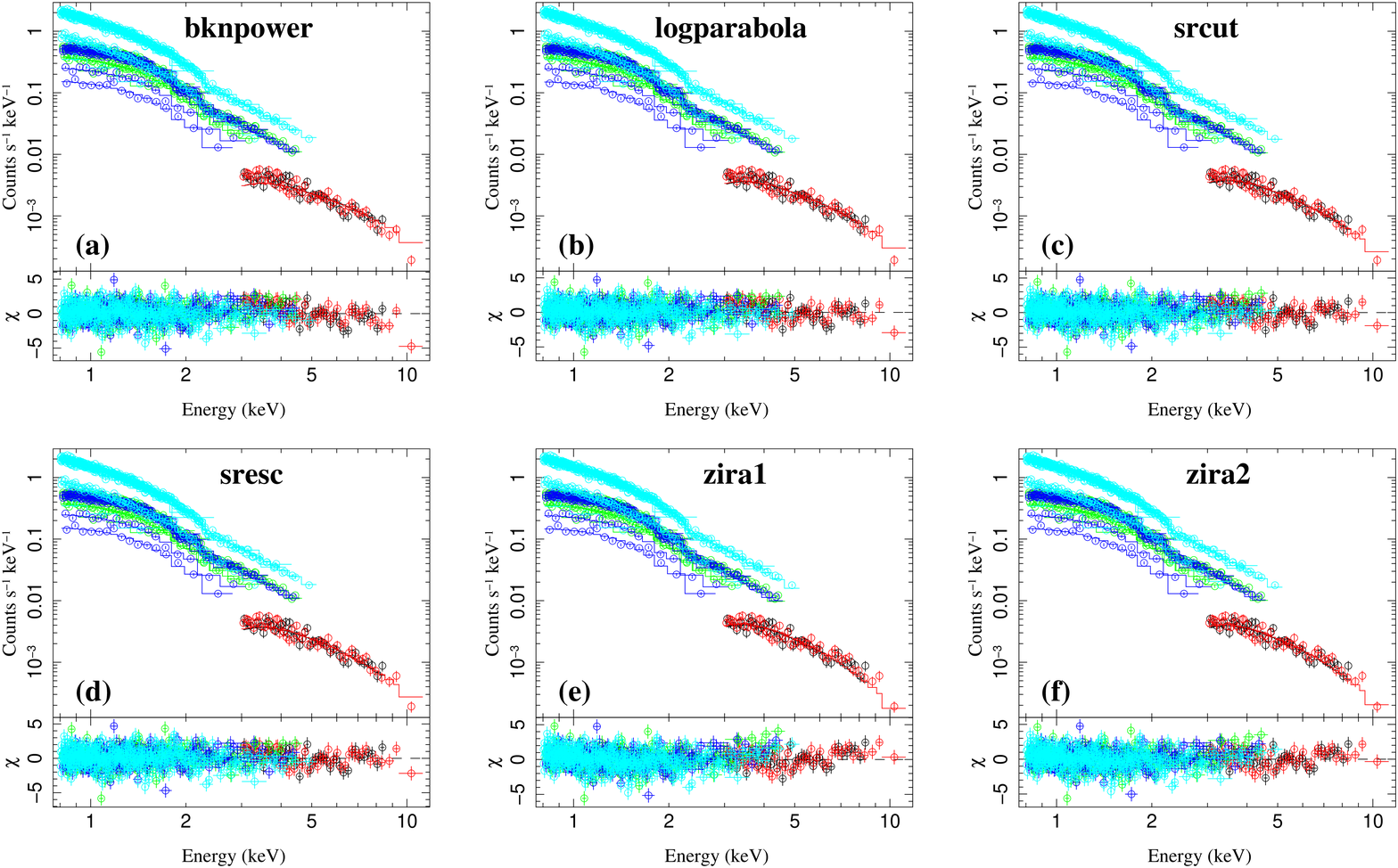,width=1.0\textwidth,angle=0, clip=}
\vspace{-0.3in}
\caption{continued, for region ``7'' in the NE limb.}
\end{center}
\end{figure*}
\addtocounter{figure}{-1}
\begin{figure*}
\begin{center}
\epsfig{figure=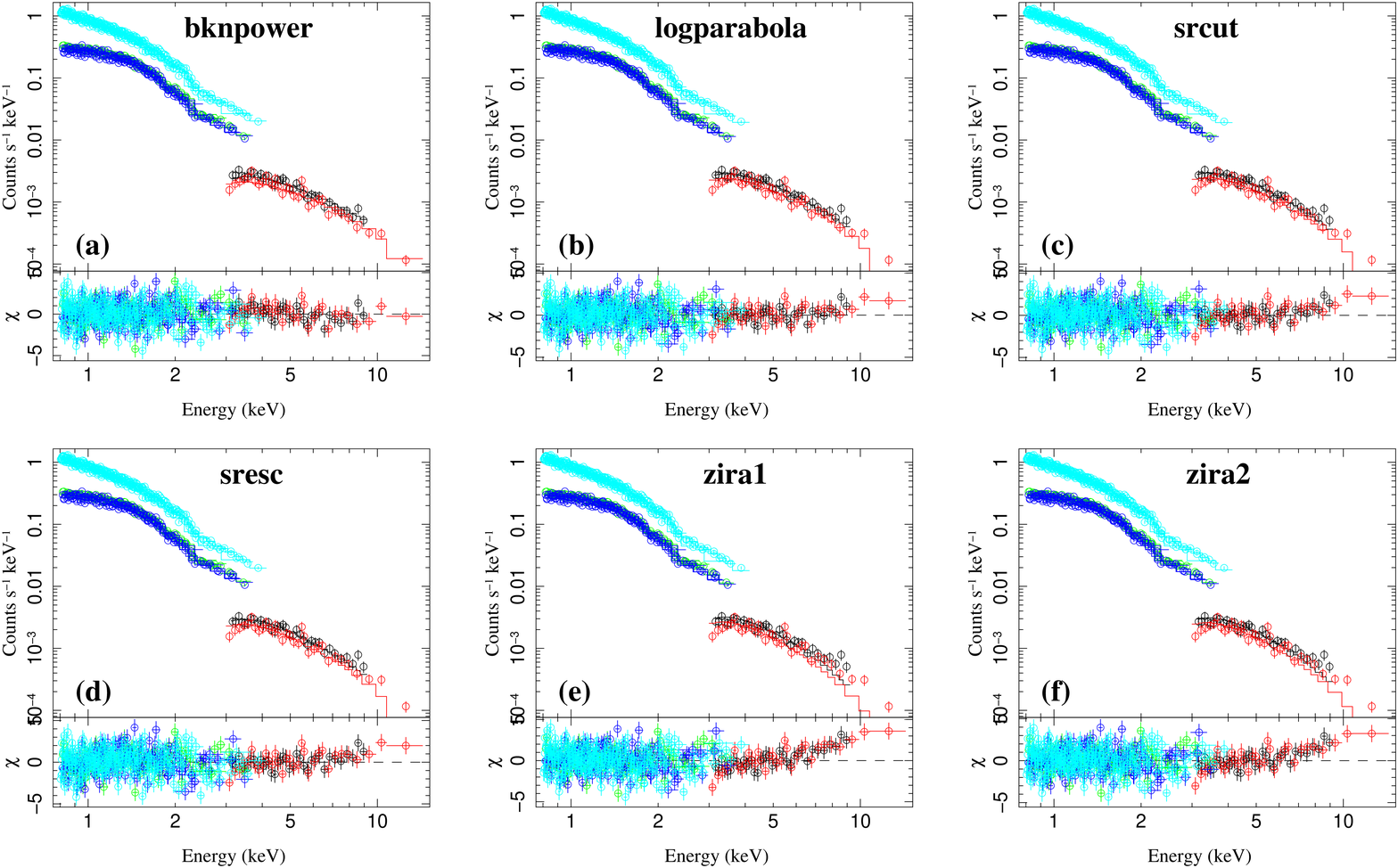,width=1.0\textwidth,angle=0, clip=}
\vspace{-0.3in}
\caption{continued, for region ``1'' in the SW limb.}
\end{center}
\end{figure*}
\addtocounter{figure}{-1}
\begin{figure*}
\begin{center}
\epsfig{figure=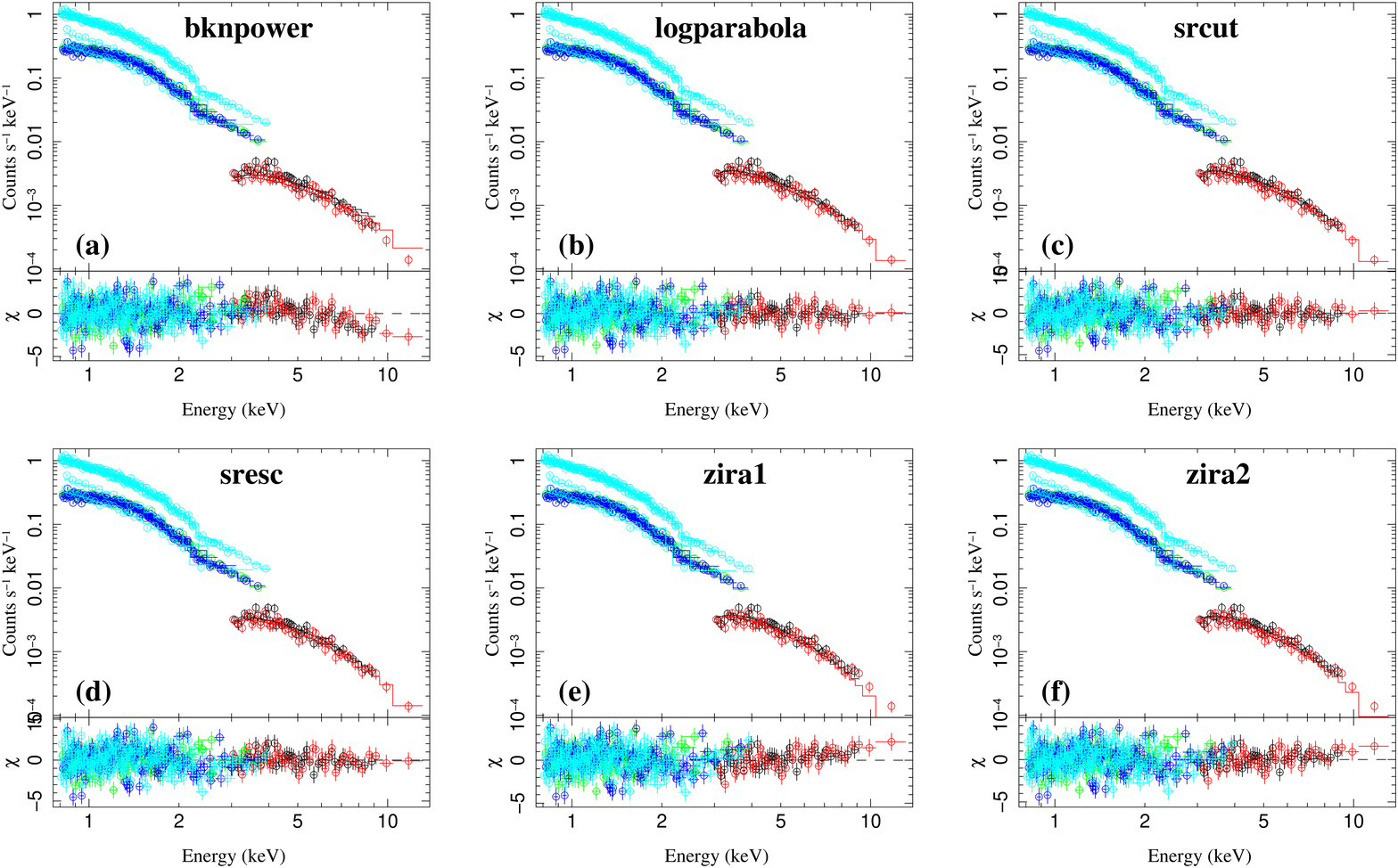,width=1.0\textwidth,angle=0, clip=}
\vspace{-0.3in}
\caption{continued, for region ``2'' in the SW limb.}
\end{center}
\end{figure*}
\addtocounter{figure}{-1}
\begin{figure*}
\begin{center}
\epsfig{figure=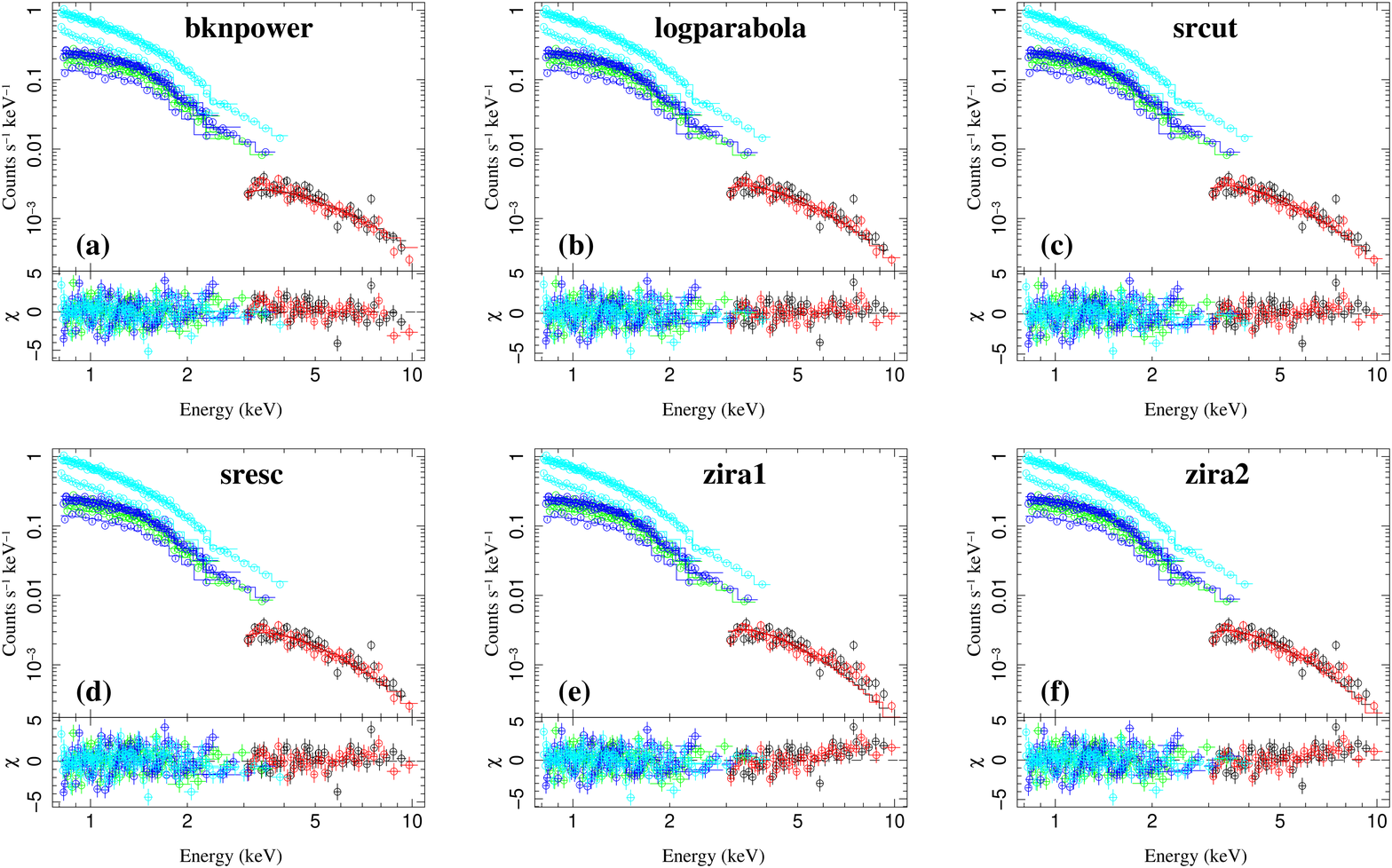,width=1.0\textwidth,angle=0, clip=}
\vspace{-0.3in}
\caption{continued, for region ``3'' in the SW limb.}
\end{center}
\end{figure*}
\addtocounter{figure}{-1}
\begin{figure*}
\begin{center}
\epsfig{figure=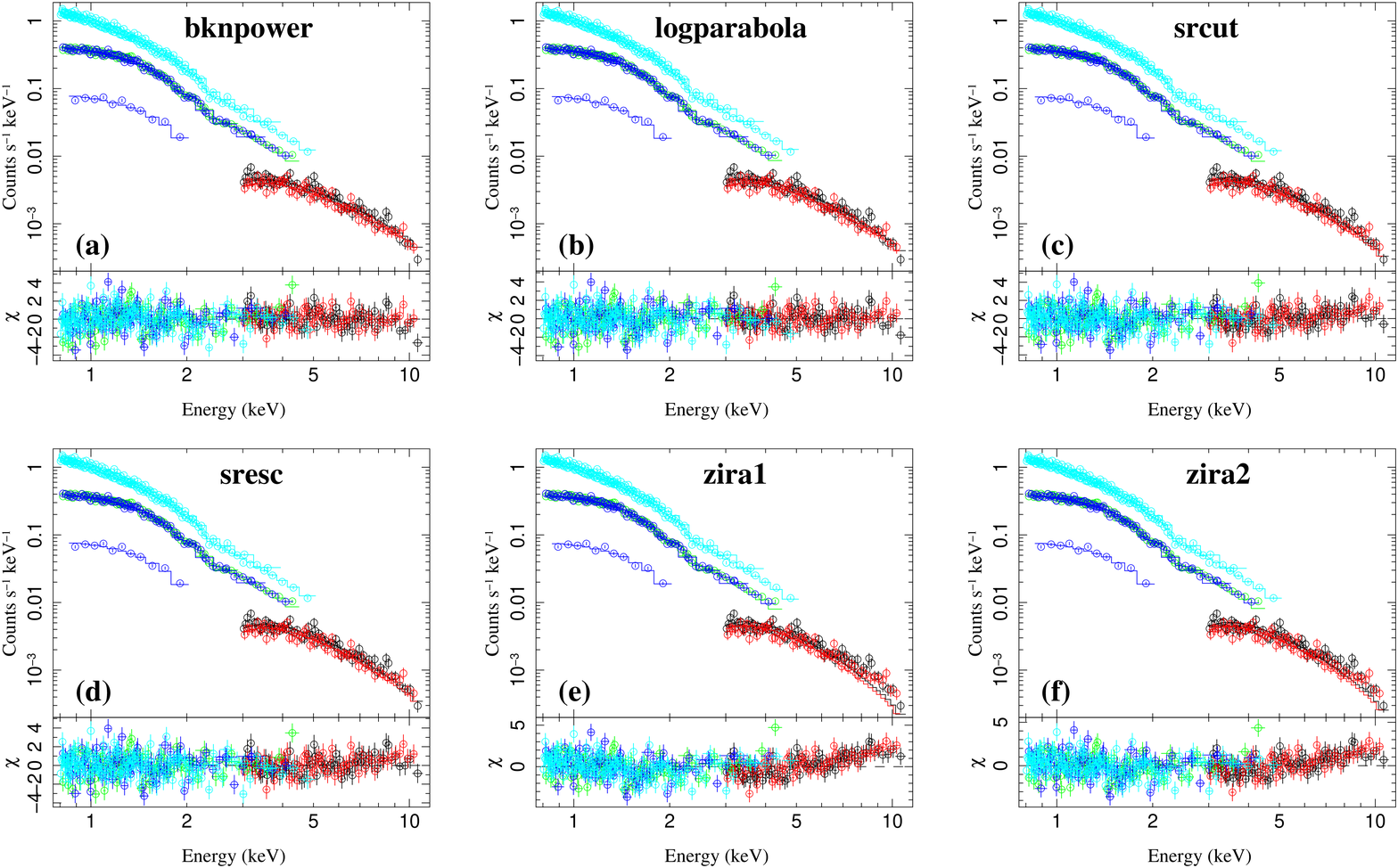,width=1.0\textwidth,angle=0, clip=}
\vspace{-0.3in}
\caption{continued, for region ``4'' in the SW limb.}
\end{center}
\end{figure*}
\addtocounter{figure}{-1}
\begin{figure*}
\begin{center}
\epsfig{figure=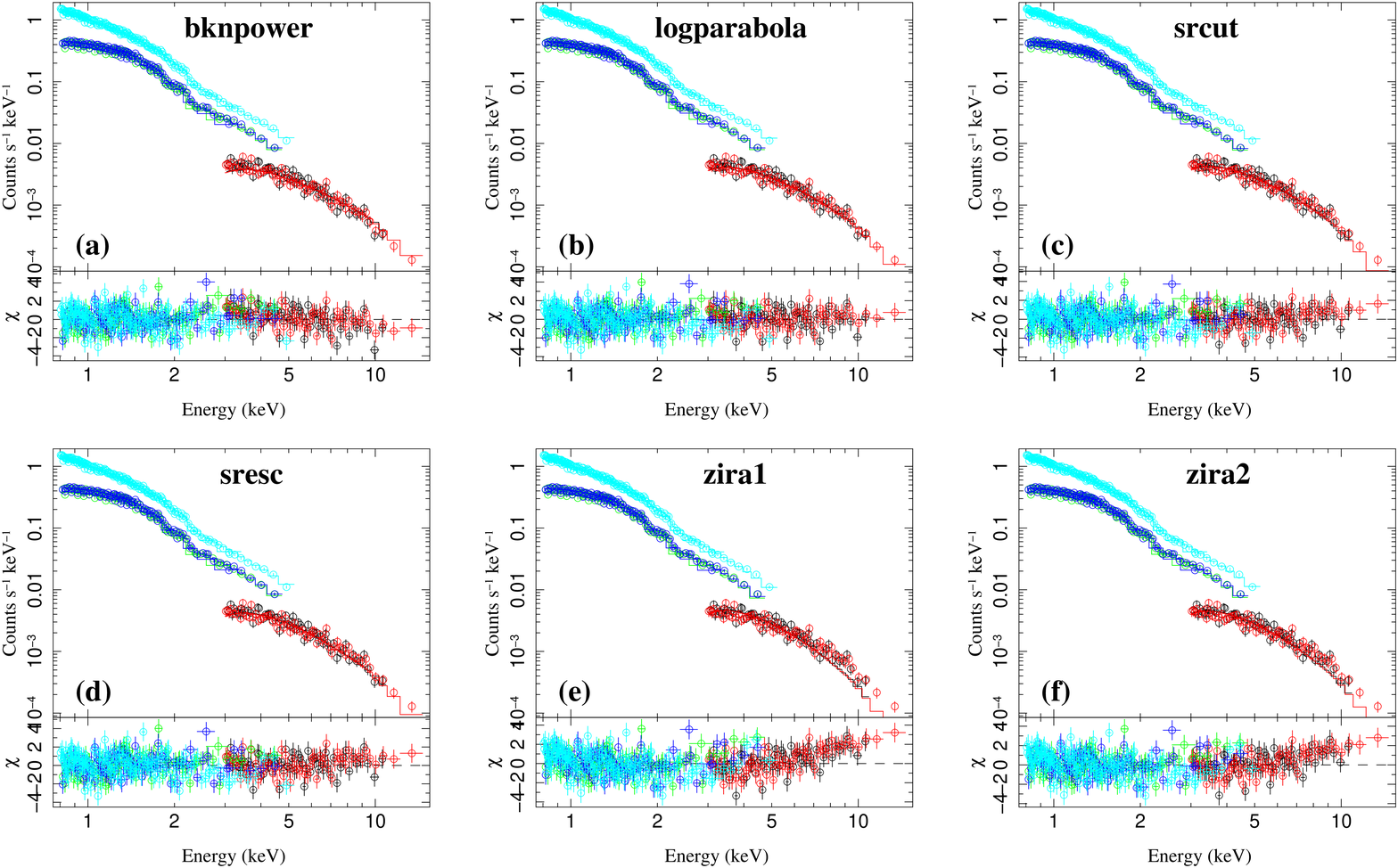,width=1.0\textwidth,angle=0, clip=}
\vspace{-0.3in}
\caption{continued, for region ``5'' in the SW limb.}
\end{center}
\end{figure*}
\addtocounter{figure}{-1}
\begin{figure*}
\begin{center}
\epsfig{figure=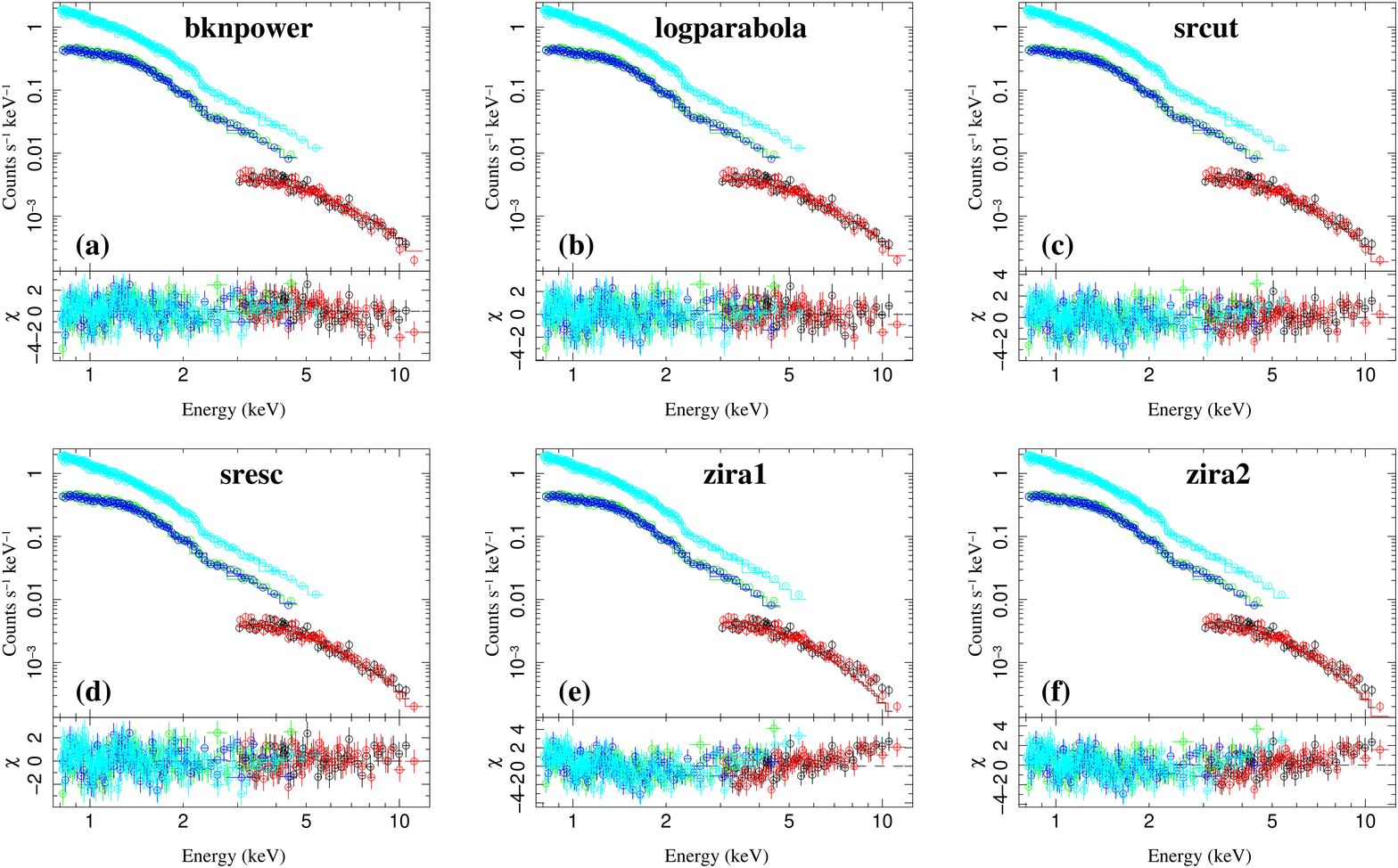,width=1.0\textwidth,angle=0, clip=}
\vspace{-0.3in}
\caption{continued, for region ``6'' in the SW limb.}
\end{center}
\end{figure*}
\addtocounter{figure}{-1}
\begin{figure*}
\begin{center}
\epsfig{figure=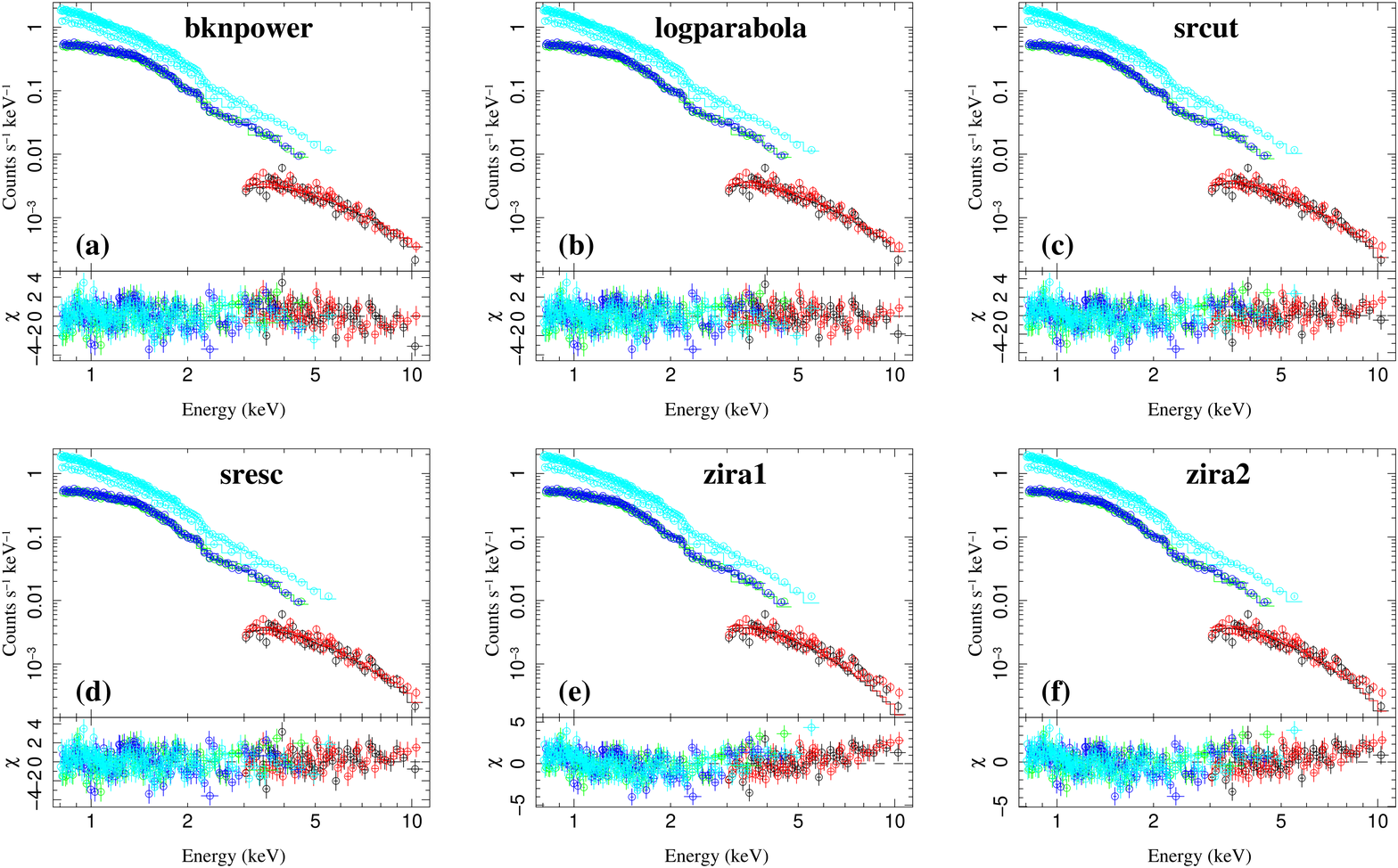,width=1.0\textwidth,angle=0, clip=}
\vspace{-0.3in}
\caption{continued, for region ``7'' in the SW limb.}
\end{center}
\end{figure*}

\end{appendices}

\end{document}